\pdfoutput=1

\documentclass[12pt,a4paper]{article}

\usepackage{ifthen} 
\newboolean{pdflatex}
\setboolean{pdflatex}{true} 

\newboolean{articletitles}
\setboolean{articletitles}{true} 

\newboolean{uprightparticles}
\setboolean{uprightparticles}{false} 

\newboolean{inbibliography}
\setboolean{inbibliography}{false} 

\usepackage{microtype}
\usepackage{lineno}  
\usepackage{xspace} 
\usepackage{caption} 

\usepackage{graphicx}  
\usepackage{color}
\usepackage{colortbl}
\graphicspath{{./figs/}} 

\usepackage{amsmath} 
\usepackage{amssymb}
\usepackage{amsfonts}
\usepackage{upgreek} 

\newcommand*\patchAmsMathEnvironmentForLineno[1]{%
\expandafter\let\csname old#1\expandafter\endcsname\csname #1\endcsname
\expandafter\let\csname oldend#1\expandafter\endcsname\csname
end#1\endcsname
 \renewenvironment{#1}%
   {\linenomath\csname old#1\endcsname}%
   {\csname oldend#1\endcsname\endlinenomath}%
}
\newcommand*\patchBothAmsMathEnvironmentsForLineno[1]{%
  \patchAmsMathEnvironmentForLineno{#1}%
  \patchAmsMathEnvironmentForLineno{#1*}%
}
\AtBeginDocument{%
\patchBothAmsMathEnvironmentsForLineno{equation}%
\patchBothAmsMathEnvironmentsForLineno{align}%
\patchBothAmsMathEnvironmentsForLineno{flalign}%
\patchBothAmsMathEnvironmentsForLineno{alignat}%
\patchBothAmsMathEnvironmentsForLineno{gather}%
\patchBothAmsMathEnvironmentsForLineno{multline}%
}

\usepackage{hyperref}    
\usepackage[all]{hypcap} 

\def\lhcb {\mbox{LHCb}\xspace}

\def\MagUp {\mbox{\em Mag\kern -0.05em Up}\xspace}

\ifthenelse{\boolean{uprightparticles}}%
{

 \def\Ppi         {\ensuremath{\uppi}\xspace}

 \def\Ppsi        {\ensuremath{\uppsi}\xspace}

 \def\PDelta      {\ensuremath{\Delta}\xspace}                 
 \def\PXi      {\ensuremath{\Xi}\xspace}                 
 \def\PLambda      {\ensuremath{\Lambda}\xspace}                 
 \def\PSigma      {\ensuremath{\Sigma}\xspace}                 
 \def\POmega      {\ensuremath{\Omega}\xspace}                 
 \def\PUpsilon      {\ensuremath{\Upsilon}\xspace}                 
 
 \def\PB      {\ensuremath{\mathrm{B}}\xspace}                 
                  
 \def\PD      {\ensuremath{\mathrm{D}}\xspace}

 \def\PJ      {\ensuremath{\mathrm{J}}\xspace}                 
 \def\PK      {\ensuremath{\mathrm{K}}\xspace}

 \def\Pb      {\ensuremath{\mathrm{b}}\xspace}                 
 \def\Pc      {\ensuremath{\mathrm{c}}\xspace}

 \def\Ph      {\ensuremath{\mathrm{h}}\xspace}                 
 \def\Pi      {\ensuremath{\mathrm{i}}\xspace}

 \def\Pp      {\ensuremath{\mathrm{p}}\xspace}

 \def\Ps      {\ensuremath{\mathrm{s}}\xspace}

}
{

 \def\Ppi         {\ensuremath{\pi}\xspace}

 \def\Ppsi        {\ensuremath{\psi}\xspace}                 
                  
 \mathchardef\PDelta="7101
 \mathchardef\PXi="7104
 \mathchardef\PLambda="7103
 \mathchardef\PSigma="7106
 \mathchardef\POmega="710A
 \mathchardef\PUpsilon="7107
                  
 \def\PB      {\ensuremath{B}\xspace}                 
                  
 \def\PD      {\ensuremath{D}\xspace}

 \def\PJ      {\ensuremath{J}\xspace}                 
 \def\PK      {\ensuremath{K}\xspace}

 \def\Pb      {\ensuremath{b}\xspace}                 
 \def\Pc      {\ensuremath{c}\xspace}

 \def\Ph      {\ensuremath{h}\xspace}                 
 \def\Pi      {\ensuremath{i}\xspace}

 \def\Pp      {\ensuremath{p}\xspace}

 \def\Ps      {\ensuremath{s}\xspace}

}

\makeatletter
\ifcase \@ptsize \relax
  \newcommand{\miniscule}{\@setfontsize\miniscule{4}{5}}
\or
  \newcommand{\miniscule}{\@setfontsize\miniscule{5}{6}}
\or
  \newcommand{\miniscule}{\@setfontsize\miniscule{5}{6}}
\fi
\makeatother

\DeclareRobustCommand{\optbar}[1]{\shortstack{{\miniscule (\rule[.5ex]{1.25em}{.18mm})}
  \\ [-.7ex] $#1$}}

\def\squark    {{\ensuremath{\Ps}}\xspace}

\def\cquark    {{\ensuremath{\Pc}}\xspace}

\def\bquark    {{\ensuremath{\Pb}}\xspace}

\def\pion   {{\ensuremath{\Ppi}}\xspace}
\def\piz    {{\ensuremath{\pion^0}}\xspace}

\def\pip    {{\ensuremath{\pion^+}}\xspace}
\def\pim    {{\ensuremath{\pion^-}}\xspace}
\def\pipm   {{\ensuremath{\pion^\pm}}\xspace}
\def\pimp   {{\ensuremath{\pion^\mp}}\xspace}

\def\kaon    {{\ensuremath{\PK}}\xspace}
  \def\Kbar    {{\kern 0.2em\overline{\kern -0.2em \PK}{}}\xspace}

\def\KorKbar    {\kern 0.18em\optbar{\kern -0.18em K}{}\xspace}

\def\Kp      {{\ensuremath{\kaon^+}}\xspace}
\def\Km      {{\ensuremath{\kaon^-}}\xspace}

\def\Kmp     {{\ensuremath{\kaon^\mp}}\xspace}
\def\KS      {{\ensuremath{\kaon^0_{\rm\scriptscriptstyle S}}}\xspace}

\def\Kstar   {{\ensuremath{\kaon^*}}\xspace}

\def\Kstarp  {{\ensuremath{\kaon^{*+}}}\xspace}
\def\Kstarm  {{\ensuremath{\kaon^{*-}}}\xspace}
\def\Kstarpm {{\ensuremath{\kaon^{*\pm}}}\xspace}

  \def\Dbar    {{\kern 0.2em\overline{\kern -0.2em \PD}{}}\xspace}
\def\D       {{\ensuremath{\PD}}\xspace}

\def\DorDbar    {\kern 0.18em\optbar{\kern -0.18em D}{}\xspace}
\def\Dz      {{\ensuremath{\D^0}}\xspace}
\def\Dzb     {{\ensuremath{\Dbar{}^0}}\xspace}
\def\Dp      {{\ensuremath{\D^+}}\xspace}

\def\Dstarp  {{\ensuremath{\D^{*+}}}\xspace}

\def\Dsp     {{\ensuremath{\D^+_\squark}}\xspace}

\def\B       {{\ensuremath{\PB}}\xspace}
\def\Bbar    {{\ensuremath{\kern 0.18em\overline{\kern -0.18em \PB}{}}}\xspace}
\def\Bb      {{\ensuremath{\Bbar}}\xspace}
\def\BorBbar    {\kern 0.18em\optbar{\kern -0.18em B}{}\xspace}
\def\Bz      {{\ensuremath{\B^0}}\xspace}

\def\Bu      {{\ensuremath{\B^+}}\xspace}

\def\Bp      {{\ensuremath{\Bu}}\xspace}

\def\Bd      {{\ensuremath{\B^0}}\xspace}
\def\Bs      {{\ensuremath{\B^0_\squark}}\xspace}
\def\Bds     {{\ensuremath{\B^0_{(\squark)}}}\xspace}

\def\jpsi     {{\ensuremath{{\PJ\mskip -3mu/\mskip -2mu\Ppsi\mskip 2mu}}}\xspace}

  \def\Y#1S{\ensuremath{\PUpsilon{(#1S)}}\xspace}

\def\proton      {{\ensuremath{\Pp}}\xspace}

\def\Lz          {{\ensuremath{\PLambda}}\xspace}
\def\Lbar        {{\ensuremath{\kern 0.1em\overline{\kern -0.1em\PLambda}}}\xspace}
\def\LorLbar    {\kern 0.18em\optbar{\kern -0.18em \PLambda}{}\xspace}

\def\Lb      {{\ensuremath{\Lz^0_\bquark}}\xspace}

\def\Lc      {{\ensuremath{\Lz^+_\cquark}}\xspace}

\def\BF         {{\ensuremath{\cal B}}\xspace}

\newcommand{\decay}[2]{\ensuremath{#1\!\to #2}\xspace}         

\def\to                 {\ensuremath{\rightarrow}\xspace}

\def\CP                {{\ensuremath{C\!P}}\xspace}

\def\AT#1     {\ensuremath{A_{\mathrm{T}}^{#1}}\xspace}           

\def\C#1      {\ensuremath{\mathcal{C}_{#1}}\xspace}                       
\def\Cp#1     {\ensuremath{\mathcal{C}_{#1}^{'}}\xspace}                    
\def\Ceff#1   {\ensuremath{\mathcal{C}_{#1}^{\mathrm{(eff)}}}\xspace}        
\def\Cpeff#1  {\ensuremath{\mathcal{C}_{#1}^{'\mathrm{(eff)}}}\xspace}       
\def\Ope#1    {\ensuremath{\mathcal{O}_{#1}}\xspace}                       
\def\Opep#1   {\ensuremath{\mathcal{O}_{#1}^{'}}\xspace}                    

\newcommand{\tev}{\ifthenelse{\boolean{inbibliography}}{\ensuremath{~T\kern -0.05em eV}\xspace}{\ensuremath{\mathrm{\,Te\kern -0.1em V}}}\xspace}
\newcommand{\gev}{\ensuremath{\mathrm{\,Ge\kern -0.1em V}}\xspace}
\newcommand{\mev}{\ensuremath{\mathrm{\,Me\kern -0.1em V}}\xspace}
\newcommand{\kev}{\ensuremath{\mathrm{\,ke\kern -0.1em V}}\xspace}
\newcommand{\ev}{\ensuremath{\mathrm{\,e\kern -0.1em V}}\xspace}
\newcommand{\gevc}{\ensuremath{{\mathrm{\,Ge\kern -0.1em V\!/}c}}\xspace}
\newcommand{\mevc}{\ensuremath{{\mathrm{\,Me\kern -0.1em V\!/}c}}\xspace}
\newcommand{\gevcc}{\ensuremath{{\mathrm{\,Ge\kern -0.1em V\!/}c^2}}\xspace}
\newcommand{\gevgevcccc}{\ensuremath{{\mathrm{\,Ge\kern -0.1em V^2\!/}c^4}}\xspace}
\newcommand{\mevcc}{\ensuremath{{\mathrm{\,Me\kern -0.1em V\!/}c^2}}\xspace}

\def\mum  {\ensuremath{{\,\upmu\rm m}}\xspace}

\def\invfb   {\ensuremath{\mbox{\,fb}^{-1}}\xspace}

\newcommand{\stat}{\ensuremath{\mathrm{\,(stat)}}\xspace}
\newcommand{\syst}{\ensuremath{\mathrm{\,(syst)}}\xspace}

\newcommand{\chisq}{\ensuremath{\chi^2}\xspace}

\def\gsim{{~\raise.15em\hbox{$>$}\kern-.85em
          \lower.35em\hbox{$\sim$}~}\xspace}
\def\lsim{{~\raise.15em\hbox{$<$}\kern-.85em
          \lower.35em\hbox{$\sim$}~}\xspace}

\def\sPlot{\mbox{\em sPlot}}

\def\sWeights{\mbox{\em sWeights}}

\def\pt         {\mbox{$p_{\rm T}$}\xspace}
\def\et         {\mbox{$E_{\rm T}$}\xspace}

\def\evtgen     {\mbox{\textsc{EvtGen}}\xspace}

\def\geant      {\mbox{\textsc{Geant4}}\xspace}

\def\photos     {\mbox{\textsc{Photos}}\xspace}

\def\pythia     {\mbox{\textsc{Pythia}}\xspace}

\def\tell1  {TELL1\xspace}
\def\ukl1   {UKL1\xspace}

\newcommand{\eg}{\mbox{\itshape e.g.}\xspace}
\newcommand{\ie}{\mbox{\itshape i.e.}\xspace}

\ifthenelse{\boolean{uprightparticles}}%
{
 
}
{
 
}

\def\had  {\ensuremath{\Ph}\xspace}
\def\hadp  {\ensuremath{\Ph^+}\xspace}

\def\hadmp  {\ensuremath{\Ph^{\mp}}\xspace}

\def\Bdsz      {\ensuremath{\B^0_{(s)}}\xspace}

\def\Dsp  {\ensuremath{\D^{+}_{s}}\xspace}

\def\BstoKstarpKm {\decay{\Bs}{\Kstarp \Km}}

\def\BstoKstarpmKmp {\decay{\Bs}{\Kstarpm \Kmp}}

\def\BstoKstarmpip {\decay{\Bs}{\Kstarm}\pip}

\def\BdtoKstarpmKmp {\decay{\Bd}{\Kstarpm \Kmp}}

\def\BdtoKstarppim {\decay{\Bd}{\Kstarp \pim}}

\def\BstoKstarmpip {\decay{\Bs}{\Kstarm \pip}}

\newcommand{\Br}[1]{\ensuremath{\BF\left(#1\right)}\xspace}

\newcommand{\tentimes}[1]{\ensuremath{\times 10^{#1}}}

\usepackage{cite} 
\usepackage{mciteplus}

\DeclareGraphicsExtensions{.pdf,.png,.eps,.mps}

\begin{document}

\renewcommand{\thefootnote}{\fnsymbol{footnote}}
\setcounter{footnote}{1}


\begin{titlepage}
\pagenumbering{roman}

\vspace*{-1.5cm}
\centerline{\large EUROPEAN ORGANIZATION FOR NUCLEAR RESEARCH (CERN)}
\vspace*{1.5cm}
\hspace*{-0.5cm}
\begin{tabular*}{\linewidth}{lc@{\extracolsep{\fill}}r}
\ifthenelse{\boolean{pdflatex}}
{\vspace*{-2.7cm}\mbox{\!\!\!\includegraphics[width=.14\textwidth]{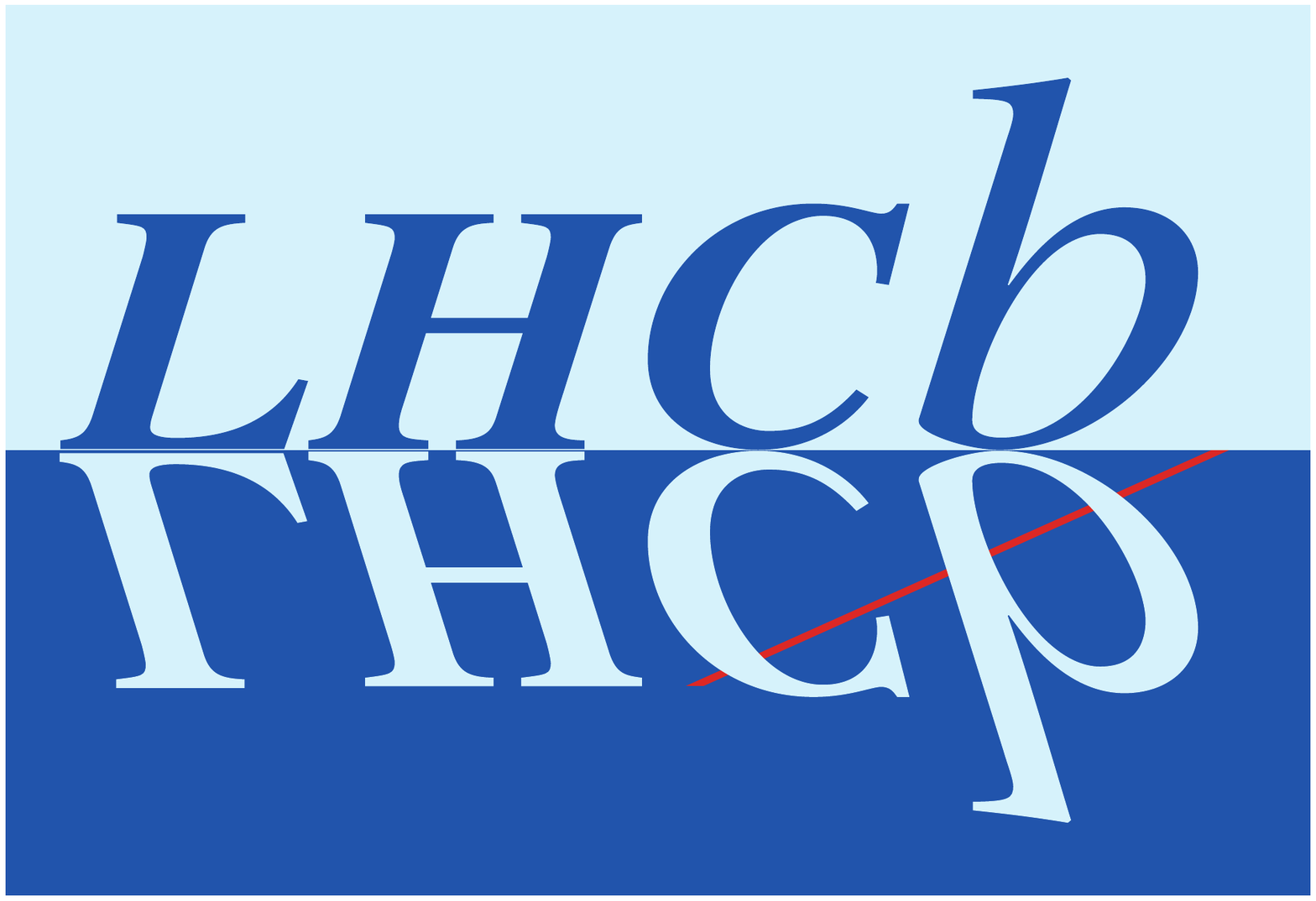}} & &}%
{\vspace*{-1.2cm}\mbox{\!\!\!\includegraphics[width=.12\textwidth]{lhcb-logo.eps}} & &}%
\\
 & & CERN-PH-EP-2014-185 \\  
 & & LHCb-PAPER-2014-043 \\  
 & & 10 October 2014 \\ 
 & & \\
\end{tabular*}

\vspace*{2.0cm}

{\bf\boldmath\huge
\begin{center}
  Observation of $B^0_s \to K^{*\pm}K^\mp$ and evidence for $B^0_s \to K^{*-}\pi^+$ decays
\end{center}
}

\vspace*{1.0cm}

\begin{center}
The LHCb collaboration\footnote{Authors are listed at the end of this paper.}
\end{center}

\vspace{\fill}

\begin{abstract}
  \noindent
  Measurements of the branching fractions of $B^0_{s} \to K^{*\pm}K^\mp$ and $B^0_{s} \to K^{*\pm}\pi^\mp$ decays are performed using a data sample corresponding to $1.0 \ {\rm fb}^{-1}$ of proton-proton collision data collected with the LHCb detector
  at a centre-of-mass energy of $7\tev$,
  where the $K^{*\pm}$ mesons are reconstructed in the $\KS\pi^\pm$ final state. 
  The first observation of the $B^0_s \to K^{*\pm}K^\mp$ decay and the first evidence for the $B^0_s \to K^{*-}\pi^+$ decay are reported with branching fractions
  \begin{eqnarray*}
  {\cal B}\left(\BstoKstarpmKmp\right) & = & \left( 12.7\pm1.9\pm1.9 \right) \times 10^{-6} \, , \\
  {\cal B}\left(\BstoKstarmpip\right)  & = & ~\left(  3.3\pm1.1\pm0.5 \right) \times 10^{-6} \, ,
  \end{eqnarray*}
  where the first uncertainties are statistical and the second are systematic.
  In addition, an upper limit of ${\cal B}\left(\BdtoKstarpmKmp\right) < 0.4 \ (0.5) \times 10^{-6}$ is set at $90\,\% \ (95\,\%)$ confidence level.
\end{abstract}

\vspace*{1.0cm}

\begin{center}
  Published in New~J.~Phys.
\end{center}

\vspace{\fill}

{\footnotesize 
\centerline{\copyright~CERN on behalf of the \lhcb collaboration, license \href{http://creativecommons.org/licenses/by/4.0/}{CC-BY-4.0}.}}
\vspace*{2mm}

\end{titlepage}

\newpage
\setcounter{page}{2}
\mbox{~}

\cleardoublepage


\renewcommand{\thefootnote}{\arabic{footnote}}
\setcounter{footnote}{0}


\pagestyle{plain} 
\setcounter{page}{1}
\pagenumbering{arabic}


\section{Introduction}
\label{sec:intro}

The Standard Model (SM) of particle physics predicts that all manifestations of \CP violation, \ie\ violation of symmetry under the combined charge conjugation and parity operation, arise due to the single complex phase that appears in the Cabibbo-Kobayashi-Maskawa (CKM) quark mixing matrix~\cite{Cabibbo:1963yz,Kobayashi:1973fv}.
Since this source is not sufficient to account for the level of the baryon asymmetry of the Universe~\cite{Riotto:1999yt}, one of the key goals of contemporary particle physics is to search for signatures of \CP violation that are not consistent with the CKM paradigm.

\begin{figure}[!b]
\centering
\includegraphics[width=0.49\textwidth]{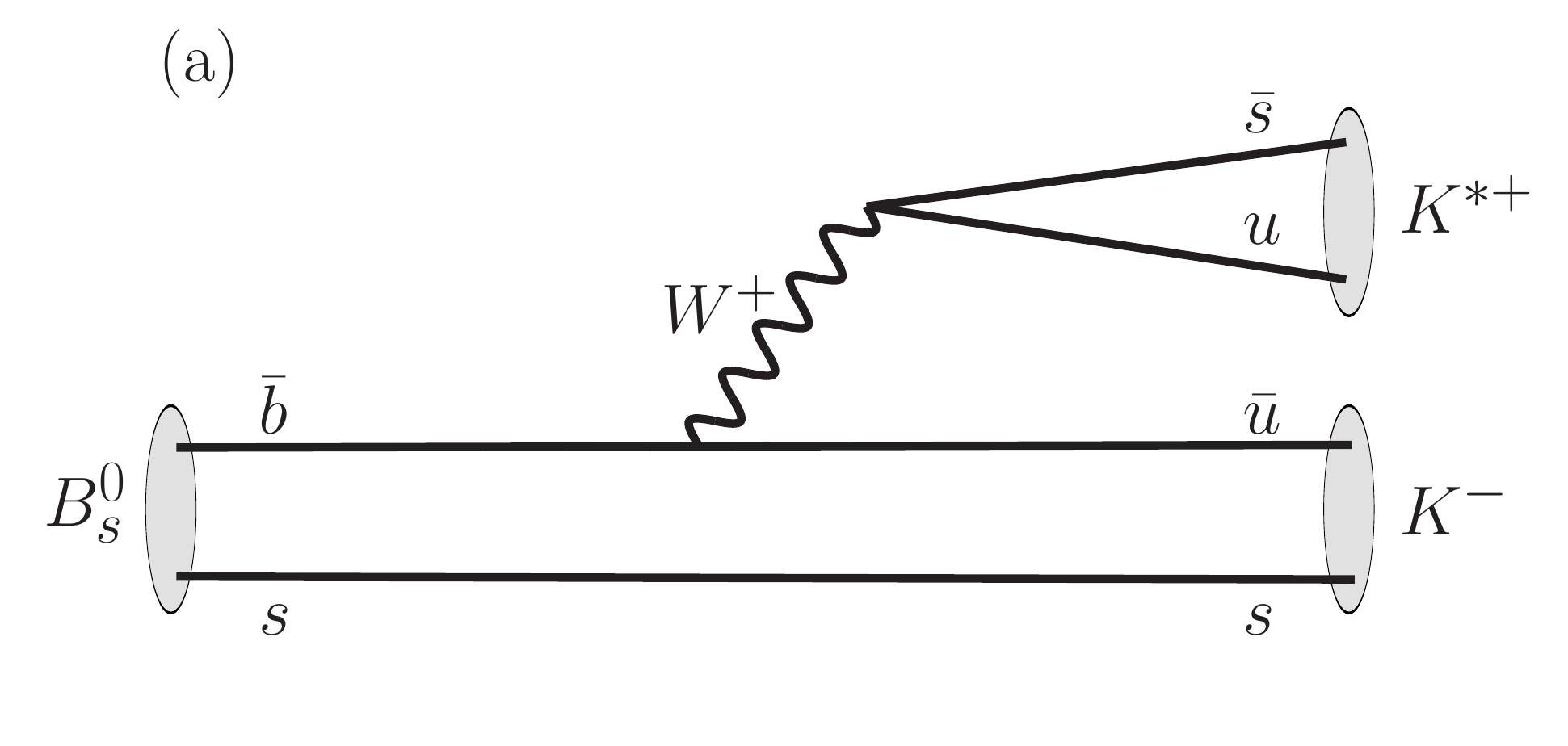}
\includegraphics[width=0.49\textwidth]{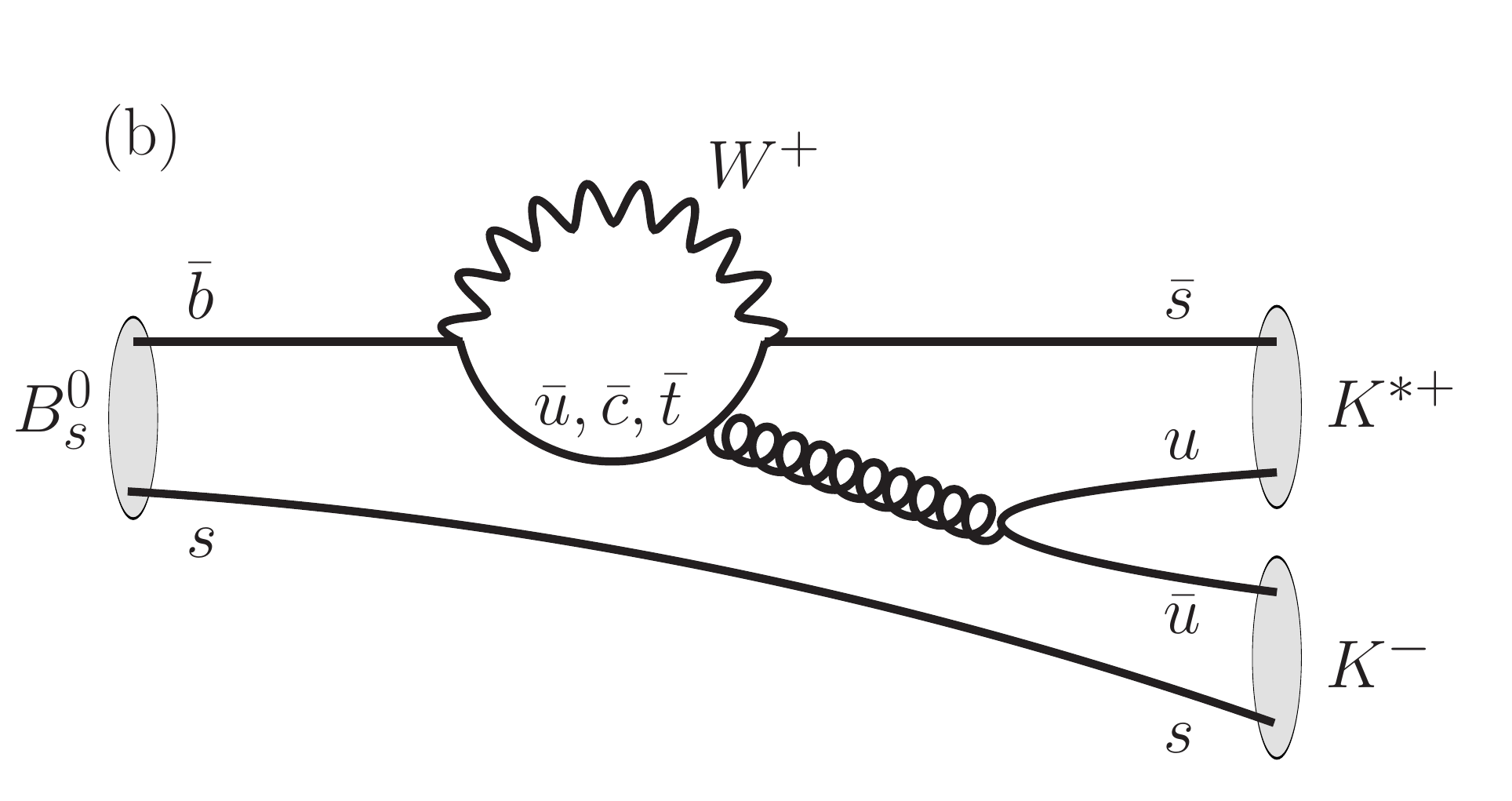}
\caption{\small
  (a) Tree and (b) loop diagrams for the decay \BstoKstarpKm.
}
\label{fig:feynman}
\end{figure}

Among the most important areas being explored in quark flavour physics is the study of $B$ meson decays to hadronic final states that do not contain charm quarks or antiquarks (hereafter referred to as ``charmless'').
As shown in Fig.~\ref{fig:feynman}, such decays have, in general, amplitudes that contain contributions from  both ``tree'' and ``loop'' diagrams (see, \eg, Ref.~\cite{Antonelli:2009ws}).
The phase differences between the two amplitudes can lead to \CP violation
and, since particles hypothesised in extensions to the SM may affect the loop diagrams, deviations from the SM predictions may occur.
Large \CP violation effects, \ie\ asymmetries of ${\cal O}(10\,\%)$ or more between the rates of $\Bb$ and $B$ meson decays to \CP conjugate final states, have been seen in $B^0 \to \Kp\pim$\cite{Lees:2012kx,Duh:2012ie,LHCb-PAPER-2013-018,Aaltonen:2014vra}, $\Bs \to \Km\pip$\cite{LHCb-PAPER-2013-018,Aaltonen:2014vra}, and $\Bp \to \pip\pim\Kp$, $\Kp\Km\Kp$, $\pip\pim\pip$ and $\Kp\Km\pip$ decays~\cite{LHCb-PAPER-2013-027,LHCb-PAPER-2013-051,LHCb-PAPER-2014-044}.
However, it is hard to be certain whether these measurements are consistent with the SM predictions due to the presence of parameters describing the hadronic interactions that are difficult to determine either theoretically or from data.

An interesting approach to control the hadronic uncertainties is to exploit amplitude analysis techniques.  
For example, by studying the distribution of kinematic configurations of $\Bd \to \KS \pip\pim$ decays across the Dalitz plot~\cite{Dalitz:1953cp}, the relative phase between the $K^{*+}\pim$ and $\KS\rho^0$ amplitudes can be determined.
This information is not accessible in studies either of two-body decays, or of the inclusive properties of three-body decays.
Consequently, it may be possible to make more sensitive tests of the SM by
studying decays to final states having contributions from intermediate states with one vector and one pseudoscalar meson (VP), rather than in those with two pseudoscalars.

Several methods to test the SM with \B meson decays to charmless VP ($K^*\pi$ and $K\rho$) states have been proposed~\cite{Ciuchini:2006kv,Ciuchini:2006st,Gronau:2006qn,Gronau:2007vr,Bediaga:2006jk,Gronau:2010dd}.
The experimental inputs needed for these methods are the magnitudes and relative phases of the decay amplitudes.  
Although the phases can only be obtained from Dalitz plot analyses of \B meson decays to final states containing one kaon and two pions, the magnitudes can be obtained from simplified approaches.
Dalitz plot analyses have been performed for the decays
$\Bp\to\Kp\pip\pim$~\cite{Garmash:2005rv,Aubert:2008bj},
$\Bd\to\KS\pip\pim$~\cite{:2008wwa,Aubert:2009me} and
$\Bd\to\Kp\pim\piz$~\cite{BABAR:2011ae}.
Decays of \B mesons to $K^*K$ final states can in principle be studied with similar methods, but the existing experimental results are less precise~\cite{Aubert:2006wu,Aubert:2007ua,Aubert:2007xb,Aubert:2008aw,delAmoSanchez:2010ur,Gaur:2013uou}.
No previous measurements of \Bs meson decays to charmless VP final states exist.
First results from the LHCb collaboration on inclusive three-body charmless \Bs decays have recently become available~\cite{LHCb-PAPER-2013-042}, but no attempt has previously been made to separate the different resonant and nonresonant contributions to their Dalitz plots.

In this paper, the first measurements of \Bs meson decays to $\Kstarm\pip$ and $\Kstarpm\Kmp$ final states and of the $\Bd \to \Kstarpm\Kmp$ rate are reported.
Throughout the remainder of the paper the symbol $\Kstar$ is used to denote the $\Kstar(892)$ resonance.
Unique charge assignments of the final state particles are specified in the expression $\Bs\to\Kstarm\pip$ because the amplitude for $\Bs \to \Kstarp\pim$ is expected to be negligibly small; however, the inclusion of charge-conjugate processes is implied throughout the paper.
The branching fractions are measured relative to that of the $\Bd\to\Kstarp\pim$ decay, which is known from previous measurements, ${\cal B}\left(\Bd\to\Kstarp\pim\right) = \left(8.5\pm0.7\right)\times 10^{-6}$~\cite{HFAG}.
Each of the relative branching fractions for $\Bs\to\Kstarpm\hadmp$, where
\had refers either to a pion or kaon, are determined as
\begin{equation}
  \frac{\Br{B^0_{s}\to\Kstarpm\hadmp}}{\Br{\Bd\to\Kstarp\pim}} = 
  \frac{f_{d}}{f_{s}} \,
  \frac{\epsilon(\Bd\to\Kstarp\pim)}{\epsilon(B^0_{s}\to\Kstarpm\hadmp)} \,
  \frac{N(B^0_{s}\to\Kstarpm\hadmp)}{N(\Bd\to\Kstarp\pim)}\, ,
  \label{eq:master-formula-Bs}
\end{equation}
while that for $\Bd\to\Kstarpm\Kmp$ is determined as
\begin{equation}
  \frac{\Br{\Bz\to\Kstarpm\Kmp}}{\Br{\Bd\to\Kstarp\pim}} = 
  \frac{\epsilon(\Bd\to\Kstarp\pim)}{\epsilon(\Bz\to\Kstarpm\Kmp)} \,
  \frac{N(\Bz\to\Kstarpm\Kmp)}{N(\Bd\to\Kstarp\pim)}\, ,
  \label{eq:master-formula-Bd}
\end{equation}
where $N$ are signal yields obtained from data, $\epsilon$ are efficiencies
obtained from simulation and corrected for known discrepancies between data
and simulation, and the ratio of fragmentation fractions $f_s/f_d = 0.259 \pm 0.015$~\cite{LHCb-PAPER-2011-018,LHCb-PAPER-2012-037,LHCb-CONF-2013-011}.
With this approach, several potentially large systematic uncertainties cancel in the ratios.
The \Kstarpm mesons are reconstructed in their decays to $\KS\pipm$ with $\KS\to\pip\pim$ and therefore the final states $\KS\pipm\hadmp$, as well as the data sample, are identical to those studied in Ref.~\cite{LHCb-PAPER-2013-042}.

Although the analysis shares several common features to that of the previous publication~\cite{LHCb-PAPER-2013-042}, the selection is optimised independently based on the expected level of background within the allowed $\KS\pipm$ mass window.
The data sample used is too small for a detailed Dalitz plot analysis, and therefore only branching fractions are measured.
The fit used to distinguish signal from background is an unbinned maximum likelihood fit in the two dimensions of $B$ candidate and $K^*$ candidate invariant masses.
This approach allows the resonant $\B\to\Kstarpm\hadmp$ decay to be separated from other \B meson decays to the $\KS\pipm\hadmp$ final state.
It does not, however, account for interference effects between the
$\Kstarpm\hadmp$ component and other amplitudes contributing to the
Dalitz plot; possible biases due to interference are considered as a source of systematic uncertainty.

\section{The LHCb detector}
\label{sec:detector}

The analysis is based on a data sample corresponding to an integrated luminosity of $1.0 \invfb$ of $pp$ collisions at a centre-of-mass energy of $7 \tev$ recorded with the \lhcb detector at CERN.
The \lhcb detector~\cite{Alves:2008zz} is a single-arm forward
spectrometer covering the \mbox{pseudorapidity} range $2<\eta <5$,
designed for the study of particles containing \bquark or \cquark
quarks. The detector includes a high-precision tracking system
consisting of a silicon-strip vertex detector (VELO)~\cite{LHCb-DP-2014-001}
surrounding the $pp$ interaction region, a large-area silicon-strip detector
located upstream of a dipole magnet with a bending power of about
$4{\rm\,Tm}$, and three stations of silicon-strip detectors and straw
drift tubes~\cite{LHCb-DP-2013-003} placed downstream.
The tracking system provides a momentum measurement with
relative uncertainty that varies from 0.4\,\% at low momentum to 0.6\,\% at 100\gevc.
The minimum distance of a track to a primary vertex, the impact parameter, is
measured with resolution of $20\mum$ for tracks with large momentum transverse to the beamline (\pt). 
Different types of charged hadrons are distinguished using information
from two ring-imaging Cherenkov detectors~\cite{LHCb-DP-2012-003}. 
Photon, electron and
hadron candidates are identified by a calorimeter system consisting of
scintillating-pad and preshower detectors, an electromagnetic
calorimeter and a hadronic calorimeter. Muons are identified by a
system composed of alternating layers of iron and multiwire
proportional chambers~\cite{LHCb-DP-2012-002}.

The trigger~\cite{LHCb-DP-2012-004} consists of hardware and software stages.
The hadron trigger at the hardware stage requires that there is at least one particle with transverse energy $\et > 3.5 \gev$.
Events containing candidate signal decays are required to have been triggered at the hardware level in one of two ways. 
Events in the first category are triggered by particles from candidate signal
decays that have an associated calorimeter energy deposit above the threshold, while those in the second category are triggered independently of the particles associated with the signal decay.
Events that do not fall into either of these categories are not used in the subsequent analysis.
The software trigger requires a two-, three- or four-track secondary vertex with a large sum of the \pt of the tracks and a significant displacement from the primary $pp$ interaction vertices~(PVs). 
A multivariate algorithm~\cite{BBDT} is used for the identification of secondary vertices consistent with the decay of a \bquark hadron.

Simulated events are used to study the detector response to signal decays and to investigate potential sources of background.
In the simulation, $pp$ collisions are generated using \pythia~\cite{Sjostrand:2006za} with a specific \lhcb configuration~\cite{LHCb-PROC-2010-056}.  
Decays of hadronic particles are described by \evtgen~\cite{Lange:2001uf}, in which final state radiation is generated using \photos~\cite{Golonka:2005pn}. 
The interaction of the generated particles with the detector and its
response are implemented using the \geant toolkit~\cite{Allison:2006ve, *Agostinelli:2002hh} as described in Ref.~\cite{LHCb-PROC-2011-006}.

\section{Selection requirements}
\label{sec:selection}

The trigger and preselection requirements are identical to those in Ref.~\cite{LHCb-PAPER-2013-042}.
As in that analysis, and those of other final states containing \KS mesons~\cite{LHCb-PAPER-2012-035,LHCb-PAPER-2013-015,LHCb-PAPER-2013-061,LHCb-PAPER-2014-006,LHCb-PAPER-2014-016}, candidate signal decays, \ie\ combinations of tracks that are consistent with the signal hypothesis, are separated into two categories: ``long'', where both tracks from the $\KS\to\pip\pim$ decay contain hits in the VELO, and ``downstream'', where neither does. 
Both categories have associated hits in the tracking detectors downstream of the magnet.
Since long candidates have better mass, momentum and vertex resolution, different selection requirements are imposed for the two categories.

The two tracks originating from the \B decay vertex, referred to hereafter as ``bachelor'' tracks, are required not to have associated hits in the muon system.
Backgrounds from decays with charm or charmonia in the intermediate state are vetoed by removing candidates with two-body invariant mass under the appropriate final state hypothesis within $30 \mevcc$ of the known masses~\cite{PDG2012}.
Vetoes are applied for $\jpsi\to\pip\pim$ or $\Kp\Km$, $\chi_{c0} \to \pip\pim$ or $\Kp\Km$, $\Dz\to\Km\pip$, $\pip\pim$ or $\Kp\Km$, $\Dp\to\KS\pip$ or $\KS\Kp$, $\Dsp\to\KS\pip$ or $\KS\Kp$ and $\Lc\to\KS\proton$ decays.

The largest source of potential background is from random combinations of final state particles, hereafter referred to as combinatorial background.
Signal candidates are separated from this source of background with the output of a neural network~\cite{Feindt:2006pm} that is trained and optimised separately for long and downstream candidates.
In the training, simulated $B^0_s \to K^{*\pm}K^\mp$ decays are used to
represent signal, and data from the high mass sideband of $\KS\pip\pim$ candidates are used as a background sample (the sideband is $40 < m(\KS\pip\pim)-m_{B^0} < 150 \mevcc$, where $m_{B^0}$ is the known value of the $B^0$ mass~\cite{PDG2012}).
The variables used are: the values of the impact parameter \chisq, defined
as the difference in \chisq of the associated PV with and without the
considered particle, for the bachelor tracks and the \KS and \B candidates;
the vertex fit \chisq for the \KS and \B candidates; the angle between the \B candidate flight direction and the line between the associated PV and the decay vertex; the separation between the PV and the decay vertex divided by its uncertainty; and the \B candidate \pt.
Some of these variables are transformed into their logarithms or other forms that are more appropriate for numerical handling.
The consistency of the distributions of these variables between data and
simulation is confirmed for $\Bd \to \KS\pip\pim$ decays using the \sPlot\
technique~\cite{Pivk:2004ty} with the \B candidate mass as discriminating variable.

The criteria on the outputs of the neural network are chosen to optimise the probability to observe the $B^0_s \to K^{*\pm}K^\mp$ decay with significance exceeding five standard deviations ($\sigma$)~\cite{Punzi:2003bu}.
For the optimisation, an additional requirement on the $\KS\pipm$ invariant mass, $\left| m(\KS\pipm)-m_{K^{*\pm}} \right| < 100 \mevcc$ with $m_{K^{*\pm}}$ the known $\Kstar^{\pm}$ mass, calculated with the \B and \KS candidates constrained to their known masses, is imposed to select the $\Kstar^\pm$ dominated region of the phase space.
The requirements on the neural network output give signal efficiencies exceeding 90\,\% for candidates containing long \KS candidates and exceeding 80\,\%  for candidates containing downstream \KS candidates, while approximately 95\,\% and 92\,\% of the background is removed from the two categories, respectively. 

Requirements are imposed on particle identification information, primarily from the ring-imaging Cherenkov detectors~\cite{LHCb-DP-2012-003}, to separate $K^{*\pm}\Kmp$ and $K^{*\pm}\pimp$ decays.
The criteria are chosen based on optimisation of a similar figure of merit to that used to obtain the requirement on the neural network output, and retain about 70\,\% of $K^{*\pm}\Kmp$ and about 75\,\% of $K^{*\pm}\pimp$ decays.
Candidates with tracks that are likely to be protons are rejected.
After all selection requirements are applied, below $1\,\%$ of events containing one candidate also contain a second candidate; all such candidates are retained.

\section{Determination of signal yields}
\label{sec:fit}

Candidates with masses inside the fit windows of $5000 < m(\KS\pipm\hadmp) < 5500 \mevcc$ and $650 < m(\KS\pipm) < 1200 \mevcc$ are used to perform extended unbinned maximum likelihood fits to determine the signal yields.
In these fits, signal decays are separated from several categories of
background by exploiting their distributions in both $m(\KS\pipm\hadmp)$ and $m(\KS\pipm)$.
The mass of the $\KS\pipm\hadmp$ combination is calculated assigning either
the kaon or pion mass to \hadmp according to the outcome of the particle identification requirement.
A single simultaneous fit to both long and downstream candidates is performed.
Separate fits are performed for $\Kstarpm\Kmp$ and $\Kstarpm\pimp$ candidates.

In addition to the signal components and combinatorial background, candidates can originate from several other \bquark hadron decays.
Potential sources include: decays of \Bd and \Bs mesons to $\KS\pipm\hadmp$ final states without an intermediate \Kstar state (referred to as ``nonresonant'');
misidentified $\Bdsz\to\Kstarpm\hadmp$ (referred to as ``cross-feed'') and $\Lb\to\Kstarm\proton$ decays;
decays of \B mesons to charmless final states with an additional unreconstructed pion;
and $\Bp\to\Dzb\hadp$, $\Dzb\to\KS\pip\pim$ decays where the additional pion is not reconstructed.
Where branching fraction measurements exist~\cite{HFAG,PDG2012,LHCb-PAPER-2013-061}, the yields of the background sources, except that for nonresonant $\Bz\to\KS\pip\pim$ decays,  are expected to be less than 10\,\% of those for $\Bs\to\Kstarpm\Kmp$.
The branching fractions of the other nonresonant decays have not been previously determined.

The fit includes components for both \Bd and \Bs signal and nonresonant components, and the sources of background listed above.
The signal components are parametrised by a Crystal Ball (CB) function~\cite{Skwarnicki:1986xj} in \B candidate mass and a relativistic Breit-Wigner (RBW) function in \Kstar candidate mass.
The peak positions and widths of the functions for the dominant contribution (\Bs for $\Kstarpm\Kmp$, \Bd for $\Kstarpm\pimp$) are allowed to vary freely in the fit.
The relative positions of the \Bd and \Bs peaks in the \B candidate mass distribution are fixed according to the known \Bd--\Bs mass difference~\cite{PDG2012}.
The tail parameters of the CB function are fixed to the values found in fits to simulated signal events, as are the relative widths of the \Bd and \Bs shapes.
Cross-feed contributions are also described by the product of CB and RBW functions with parameters determined from simulation.  
The misidentification causes a shift and a smearing of the \B candidate mass distribution and only small changes to the shape in the \Kstar candidate mass.

The \B candidate mass distributions for the nonresonant components are also
parametrised by a CB function, with peak positions and widths identical to
those of the signal components, but with different tail parameters that are fixed to values obtained from simulation.  
Within the \Kstar mass window considered in the fit, the nonresonant shape can
be approximated with a linear function.
All linear functions used in the fit are parametrised by their yield and the abscissa value at which they cross zero, and are set to zero beyond this threshold, $m_0$.
The relative yields of nonresonant and signal components are constrained to have the same value in the samples with long and downstream candidates, but this ratio is allowed to be different for \Bd and \Bs decays.

Backgrounds from other \bquark hadron decays are described non-parametrically by kernel functions~\cite{Cranmer:2000du} in the \B candidate mass and either RBW or linear functions in the \Kstar candidate mass, depending on whether or not the decay involves a $\Kstar$ resonance.
All these background shapes are determined from simulation.  
To reduce the number of free parameters in the fit to the $\Kstarpm\Kmp$ sample, the yields of the backgrounds from charmless hadronic \B meson decays with missing particles are fixed relative to the yield for the $\Bz\to\Kstarp\pim$ cross-feed component according to expectation.
The yield of the $\Bp\to\Dzb\hadp$, $\Dzb\to\KS\pip\pim$ component is determined from the fit to data.
The yield for the $\Lb\to\Kstarm\proton$ contribution is also a free parameter in the fit to $\Kstarpm\Kmp$ candidates, but is fixed to zero in the fit to $\Kstarpm\pimp$ candidates.

The combinatorial background is modelled with linear functions in both \B and \Kstar candidate mass distributions, with parameters freely varied in the fit to data except for the $m_0$ threshold in \B candidate mass, which is fixed from fits to sideband data.
For all components, the factorisation of the two-dimensional probability density functions into the product of one-dimensional functions is verified to be a good approximation using simulation and sideband data.
In total there are 20 free parameters in the fit to the $\Kstarpm\Kmp$ sample:
yields for \Bd and \Bs signals, cross-feed, \Lb, $B\to Dh$ and combinatorial backgrounds (all for both long and downstream categories); ratios of yields for the \Bd and \Bs nonresonant components; peak position and width parameters for the signal in both \B candidate and \Kstar candidate mass distributions; and parameters of the linear functions describing the combinatorial background in \Kstar candidate mass for both long and downstream categories.
The fit to the $\Kstarpm\pimp$ sample has the same number of free parameters, with the \Lb background yields replaced by charmless background yields.
The stability of both fits is confirmed using simulated pseudoexperiments.

\begin{figure}[!t]
\centering
\includegraphics[width=0.49\textwidth]{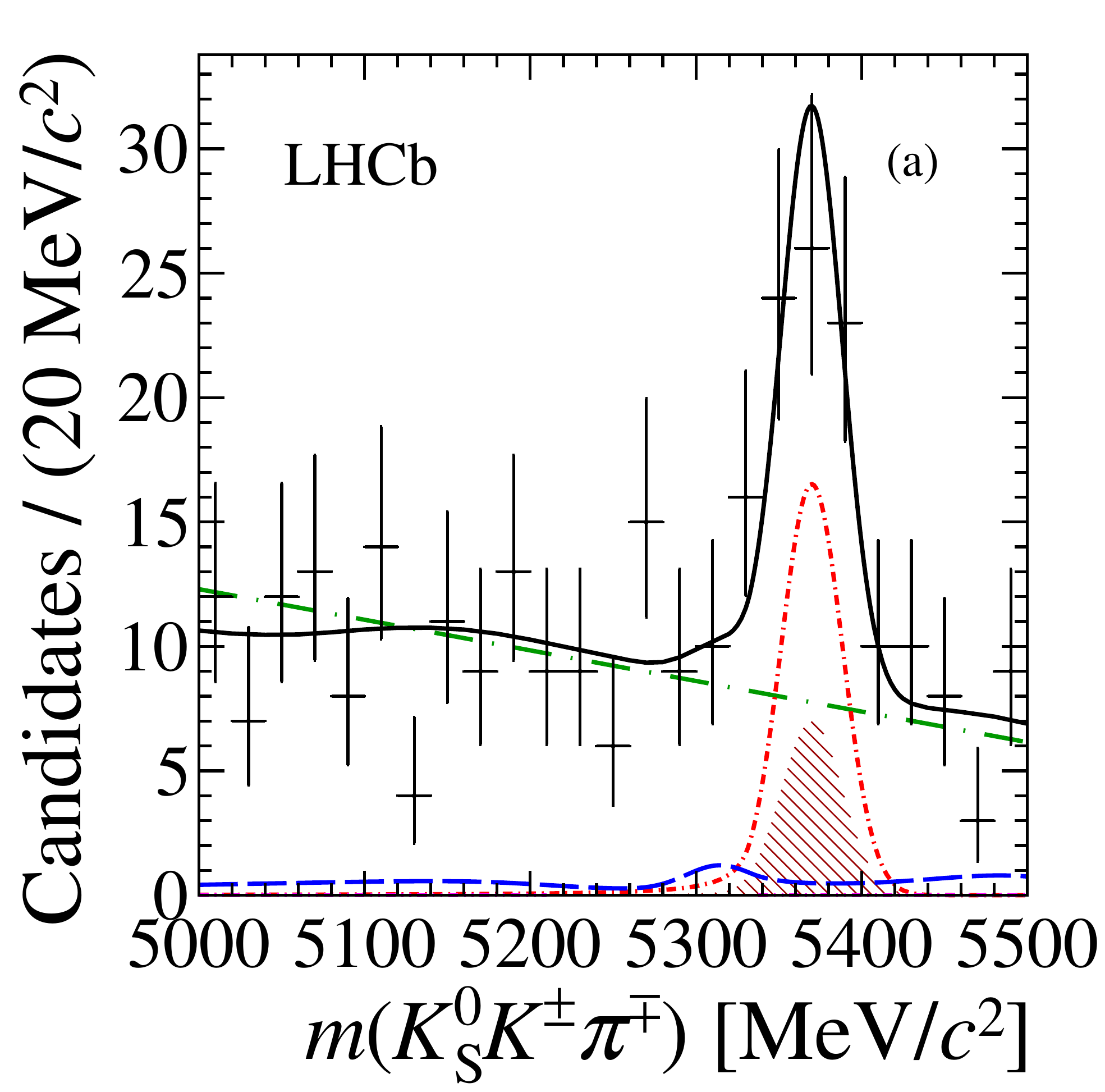}
\includegraphics[width=0.49\textwidth]{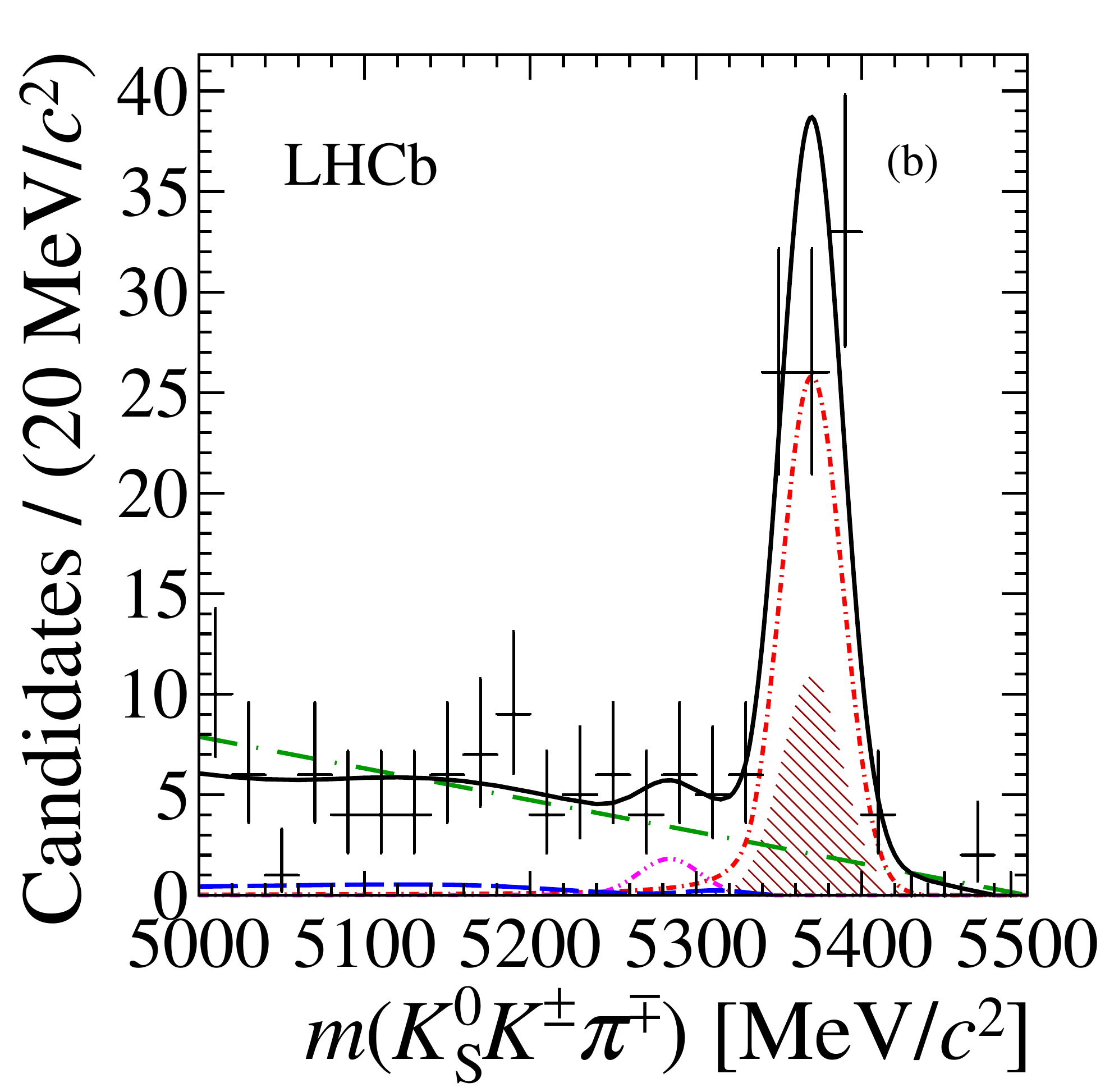} \\
\includegraphics[width=0.49\textwidth]{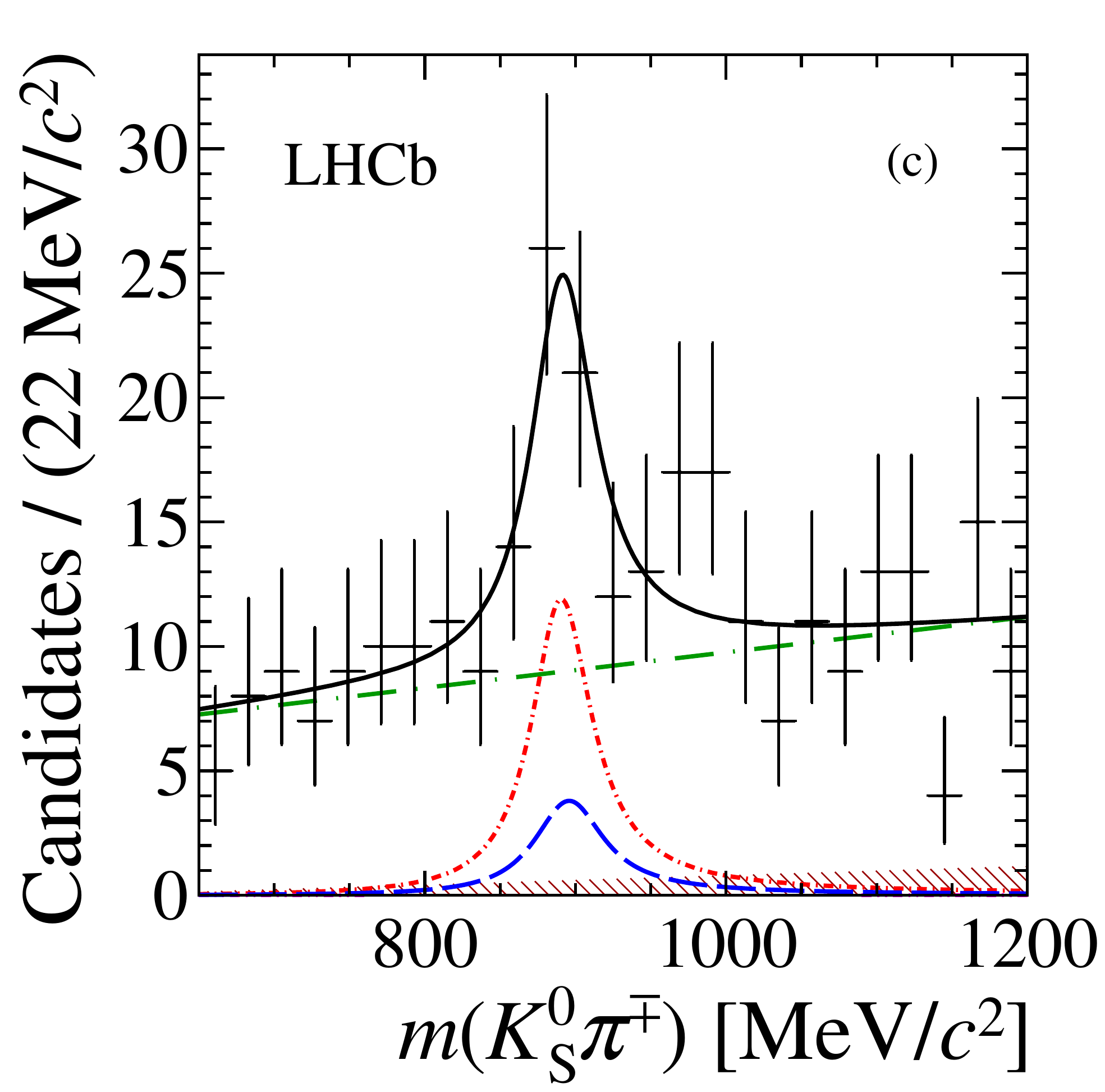}
\includegraphics[width=0.49\textwidth]{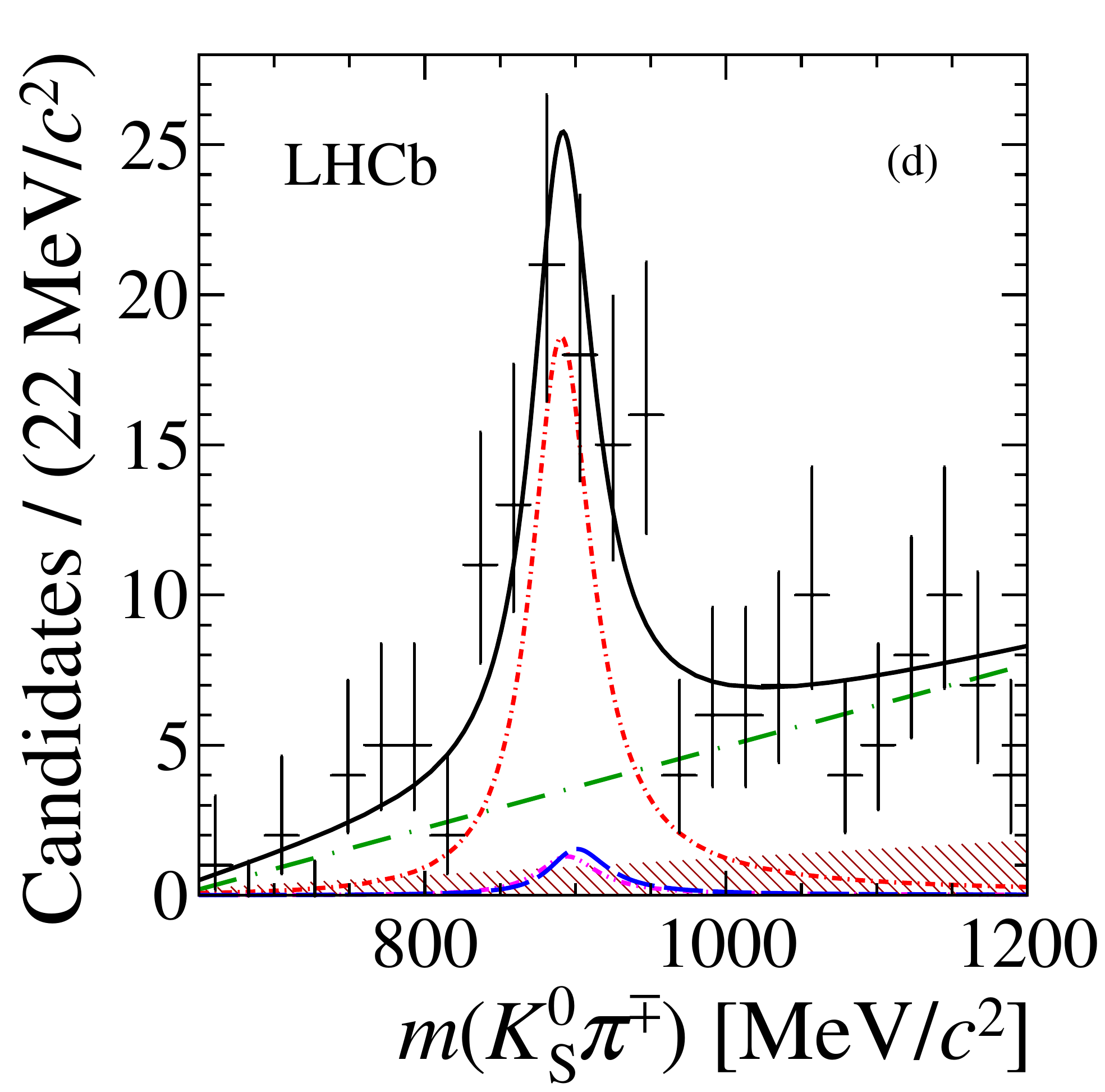}
\caption{\small
  Results of the fit to $\Kstarpm\Kmp$ candidates projected onto (a,b) \B candidate and (c,d) \Kstar candidate mass distributions, for (a,c) long and (b,d) downstream candidates.
  The total fit result (solid black line) is shown together with the data points.
  Components for the \Bd (pink dash double-dotted line) and \Bs (red dash dotted line) signals are shown
  together with the \Bs nonresonant component (dark red falling-hatched area),
  charmless partially reconstructed and cross-feed background (blue long-dashed line),
  and combinatorial background (green long-dash dotted line) components.
  The $\Bp\to\Dzb\hadp$ background component has a negative yield (consistent
  with zero) and so is not directly visible but causes the total PDF to go
  below the level of the combinatorial background on the left of the \B
  candidate mass spectrum.
}
\label{fig:KstarK}
\end{figure}

\begin{figure}[!t]
\centering
\includegraphics[width=0.49\textwidth]{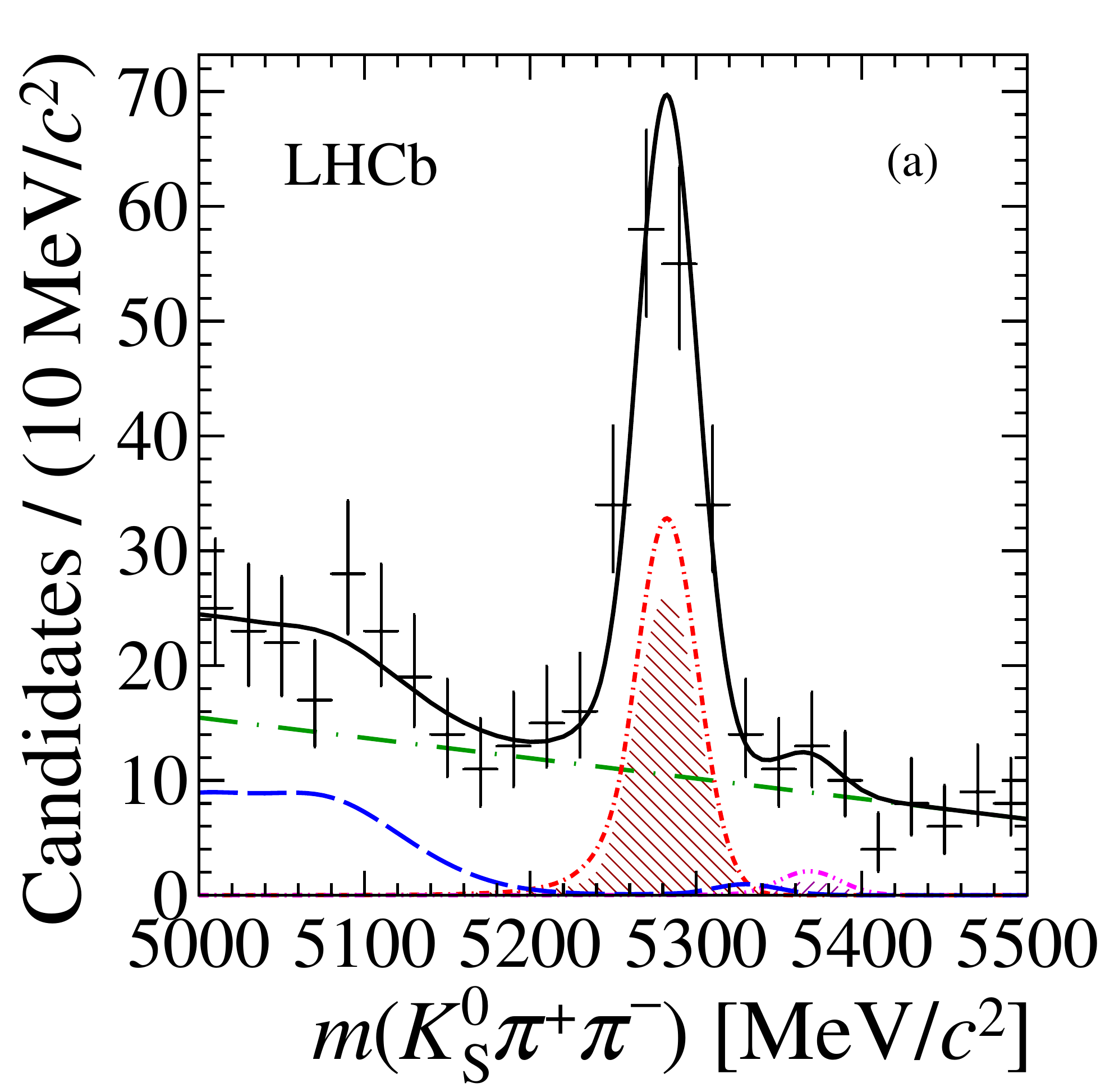}
\includegraphics[width=0.49\textwidth]{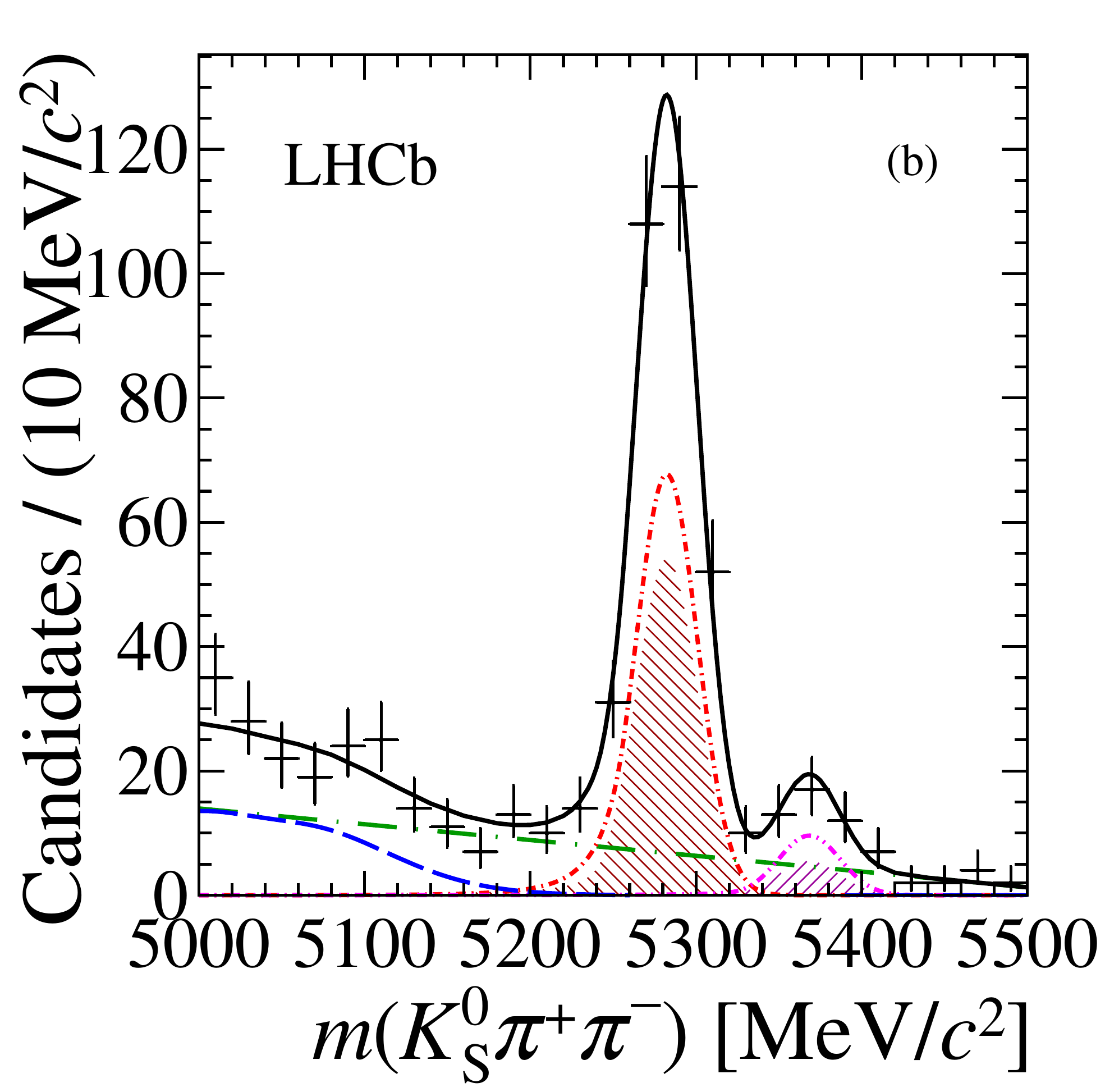} \\
\includegraphics[width=0.49\textwidth]{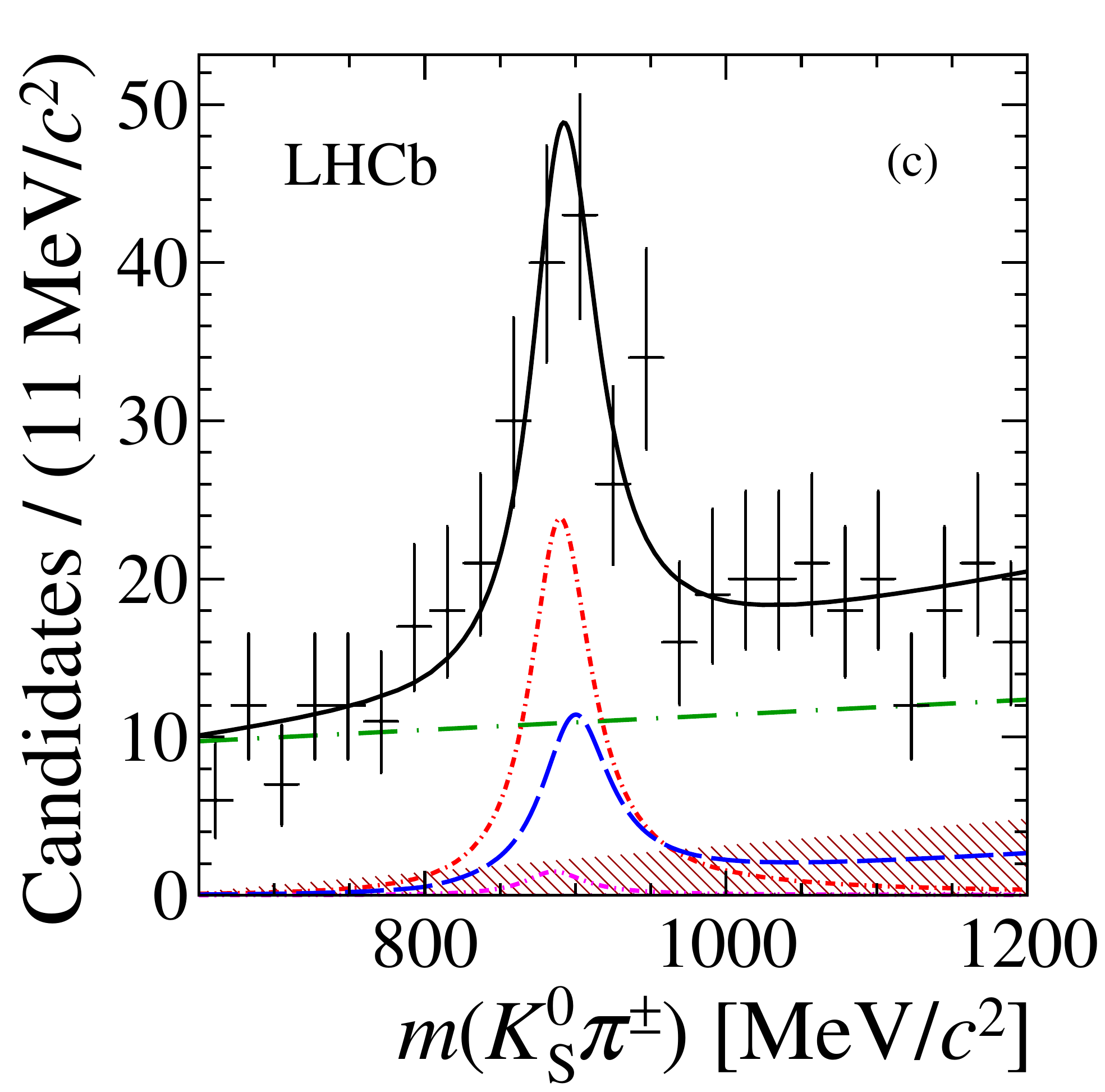}
\includegraphics[width=0.49\textwidth]{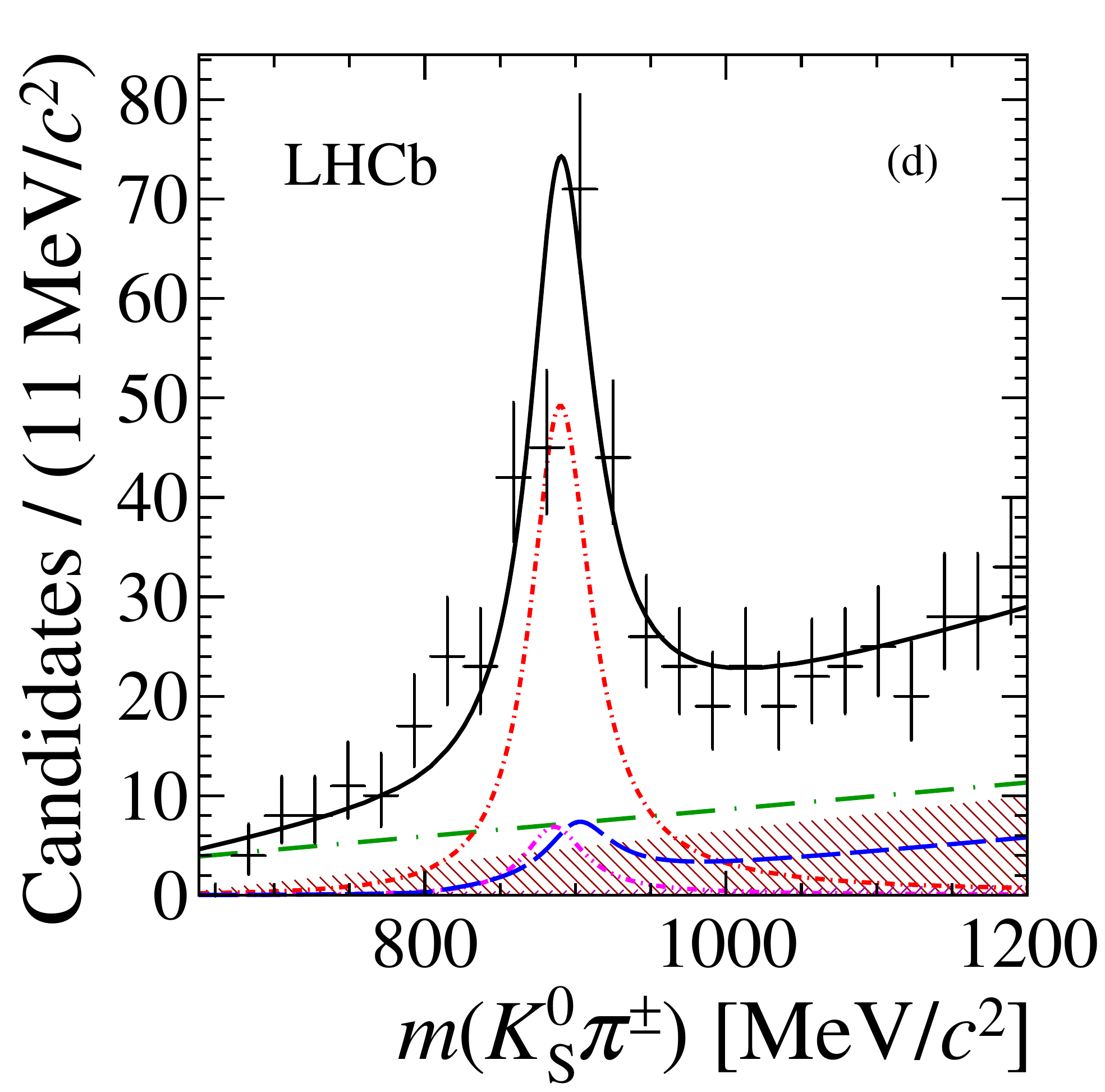}
\caption{\small
  Results of the fit to $\Kstarpm\pimp$ candidates projected onto (a,b) \B candidate and (c,d) \Kstar candidate mass distributions, for (a,c) long and (b,d) downstream candidates.
  The total fit result (black solid line) is shown together with the data points.
  Components for the \Bd (red dash dotted line) and \Bs (pink dash double-dotted line) signals are shown
  together with \Bd (dark red falling-hatched area) and \Bs (purple rising-hatched area) nonresonant components, 
  partially reconstructed and cross-feed background (blue long-dashed line),
  and combinatorial background (green long-dash-dotted line) components.
}
\label{fig:Kstarpi}
\end{figure}

The results of the fits are shown in Figs.~\ref{fig:KstarK} and~\ref{fig:Kstarpi} for the $\Kstarpm\Kmp$ and $\Kstarpm\pimp$ final states, respectively, and the signal yields are given in Table~\ref{tab:results}.
All other fit results are consistent with expectations.

\begin{table}[ h t b p ]
\centering
\caption{\small
  Yields and relative yields obtained from the fits to $\Kstarpm\Kmp$ and $\Kstarpm\pimp$ candidates.
  The relative yields of nonresonant (NR) $B^0_{(s)}$ decays are constrained to be identical in long and downstream categories.
  Only statistical uncertainties are given.
}
\label{tab:results}
\begin{tabular}{lcccc}
\hline
Yield & \multicolumn{2}{c}{\Bd} & \multicolumn{2}{c}{\Bs} \\
      & \phantom{eee}long\phantom{eee} & downstream       & \phantom{eee}long\phantom{eee} & downstream       \\
\hline
$N(\Kstarpm\Kmp)$  & $0 \pm 4$ & $\phantom{1}4 \pm 3$ & $40 \pm 8$ & $62 \pm 10$ \\
$N(\Kstarpm\pimp)$ & $80 \pm 10$ & $165 \pm 16$ & $\phantom{1}5 \pm 4$ & $23 \pm 8\phantom{1}$ \\
$N(\KS\pipm\Kmp \ \text{NR})/N(\Kstarpm\Kmp)$  & \multicolumn{2}{c}{$0.0\pm 1.0$}   & \multicolumn{2}{c}{$0.41 \pm 0.16$} \\
$N(\KS\pipm\pimp \ \text{NR})/N(\Kstarpm\pimp)$ & \multicolumn{2}{c}{$0.79\pm 0.14$} & \multicolumn{2}{c}{$0.6 \pm 0.4$} \\
\hline
\end{tabular}
\end{table}

\clearpage

\section{Systematic uncertainties}
\label{sec:systematics}

Systematic uncertainties occur due to possible imperfections in the fit model used to determine the signal yields, and due to imperfect knowledge of the efficiencies used to convert the yields to branching fraction results.
A summary of the systematic uncertainties is given in Table~\ref{tab:systematics}.

The fixed parameters in the functions describing the signal and background components are varied within their uncertainties, and the changes in the fitted yields are assigned as systematic uncertainties.  
Studies with simulated pseudoexperiments cannot exclude biases on the yields at the level of a few decays.
An uncertainty corresponding to the size of the possible bias is assigned.
The linear approximation for the shape of the nonresonant component in the \Kstar candidate mass can only be valid over a restricted range.
Therefore the mass window is varied and the change in the fitted results taken as an estimate of the corresponding uncertainty.

\begin{figure}[!t]
\centering
\includegraphics[width=0.49\textwidth]{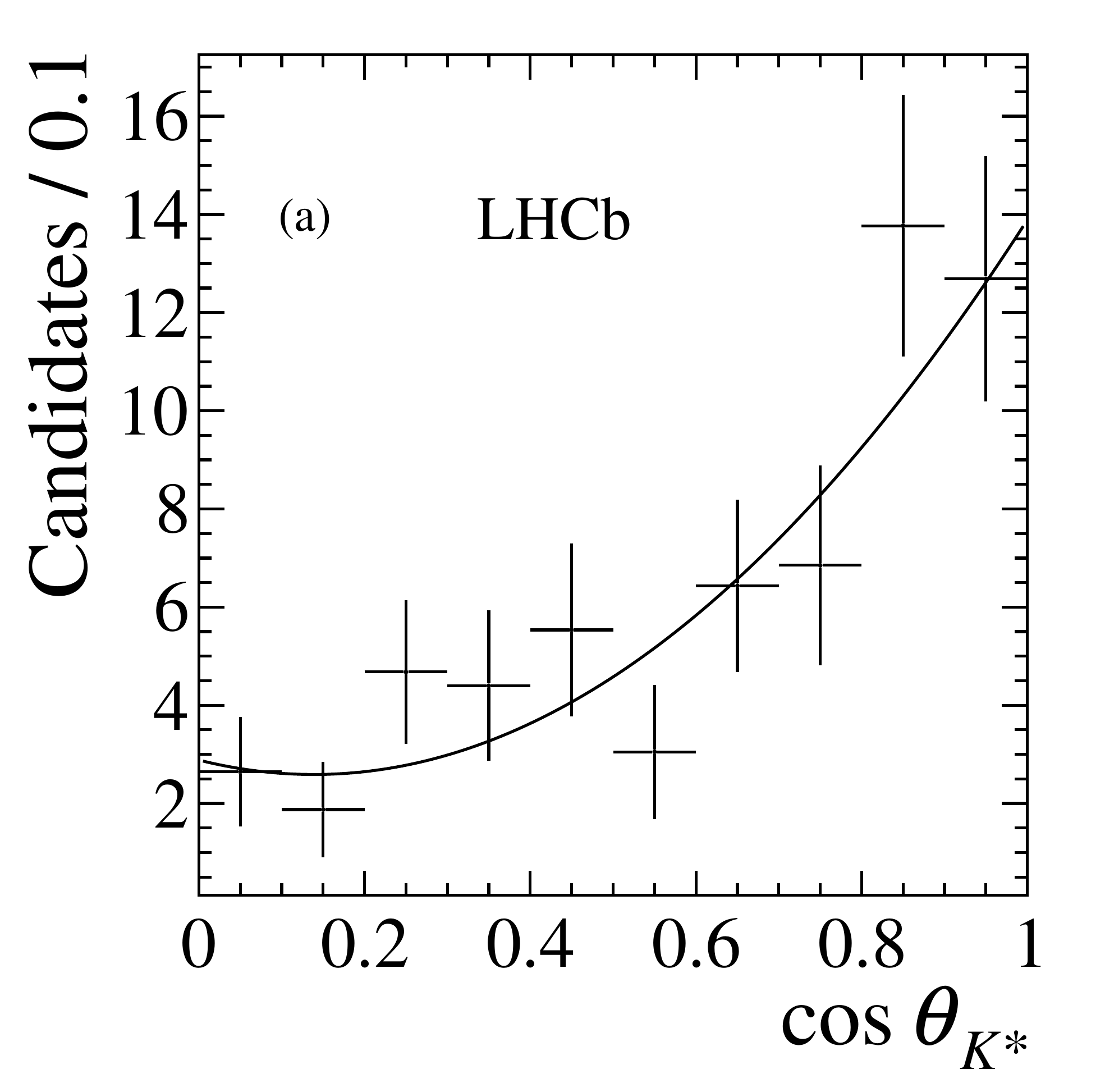}
\includegraphics[width=0.49\textwidth]{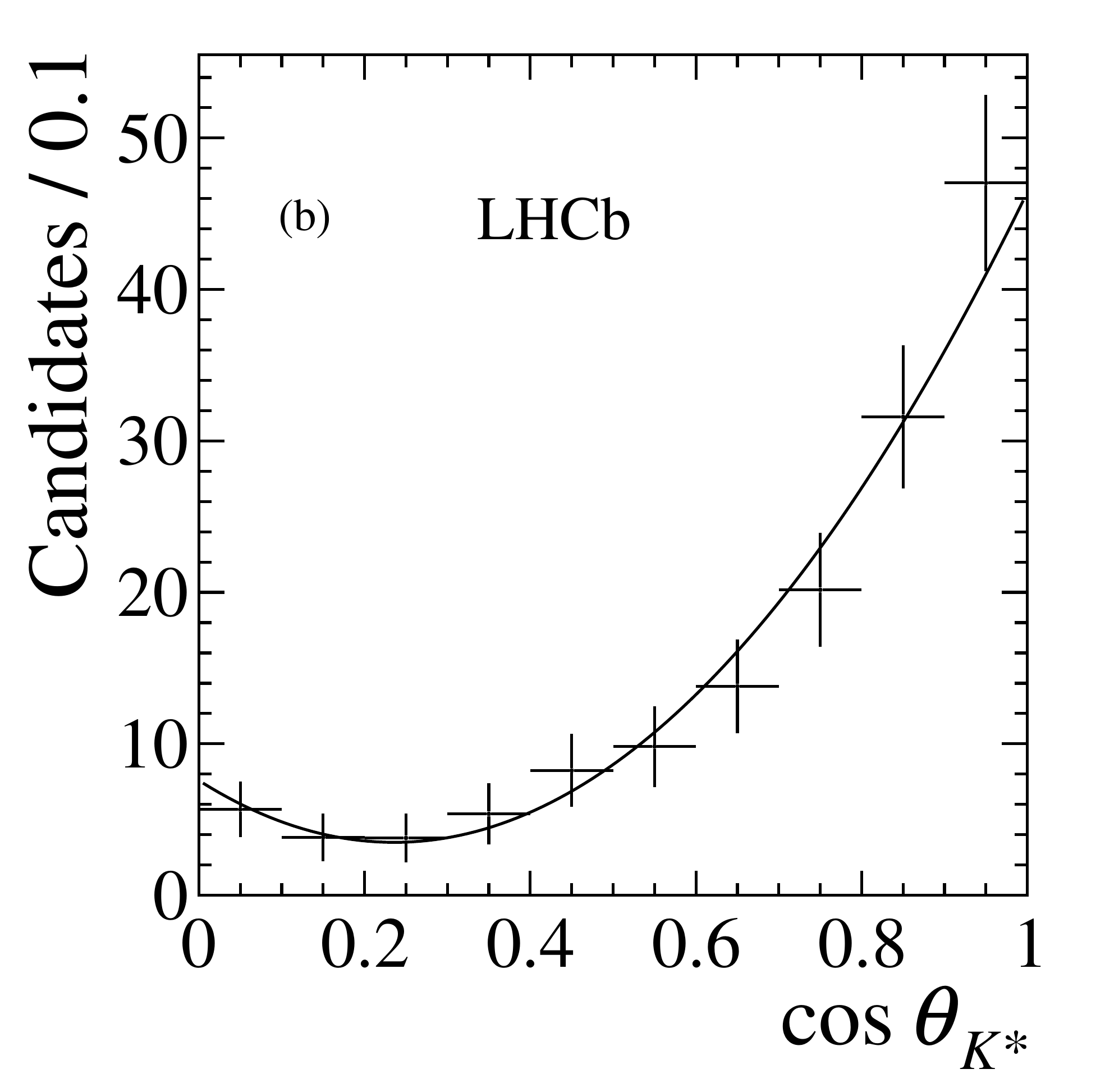}
\includegraphics[width=0.49\textwidth]{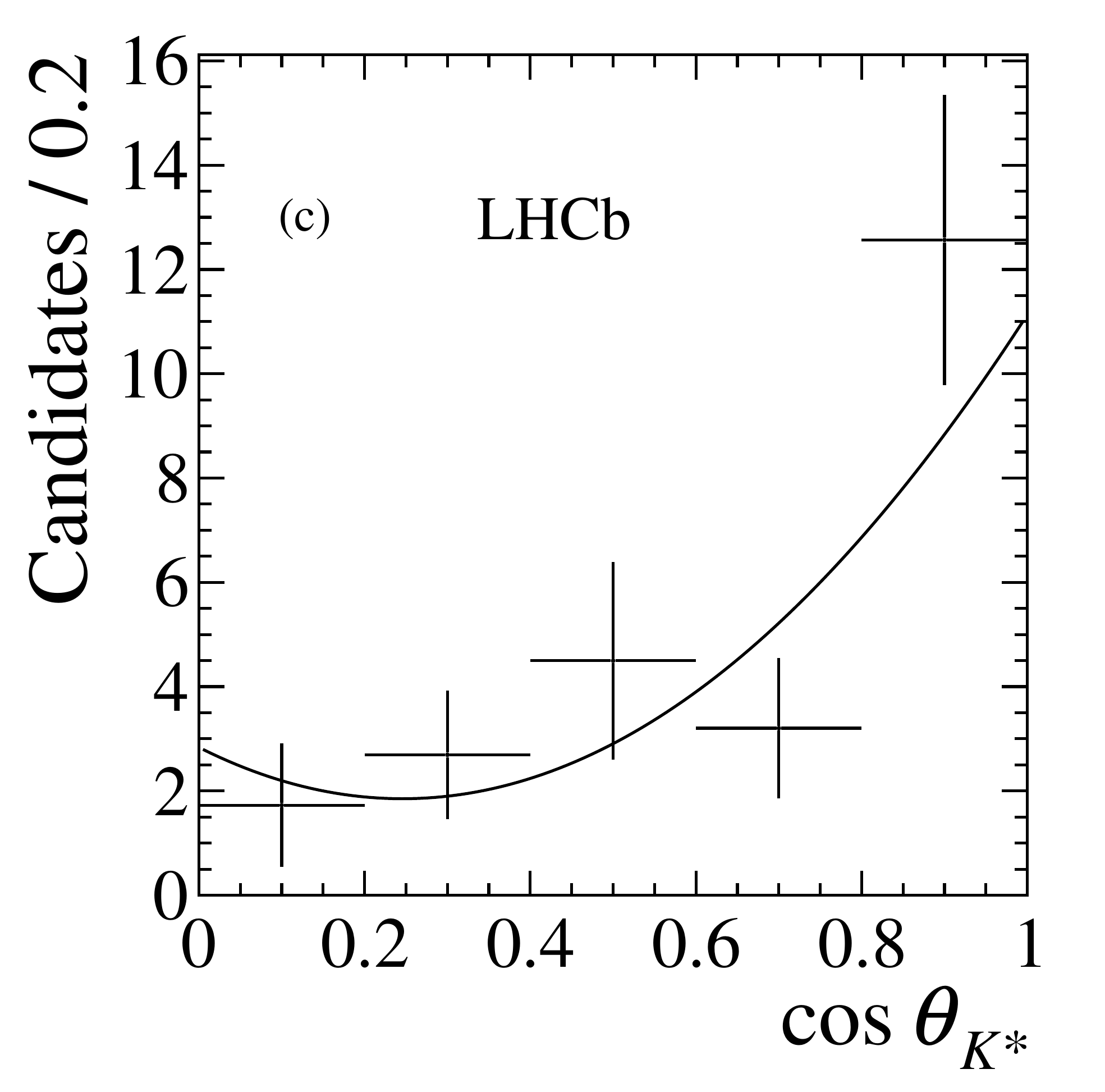}
\includegraphics[width=0.49\textwidth]{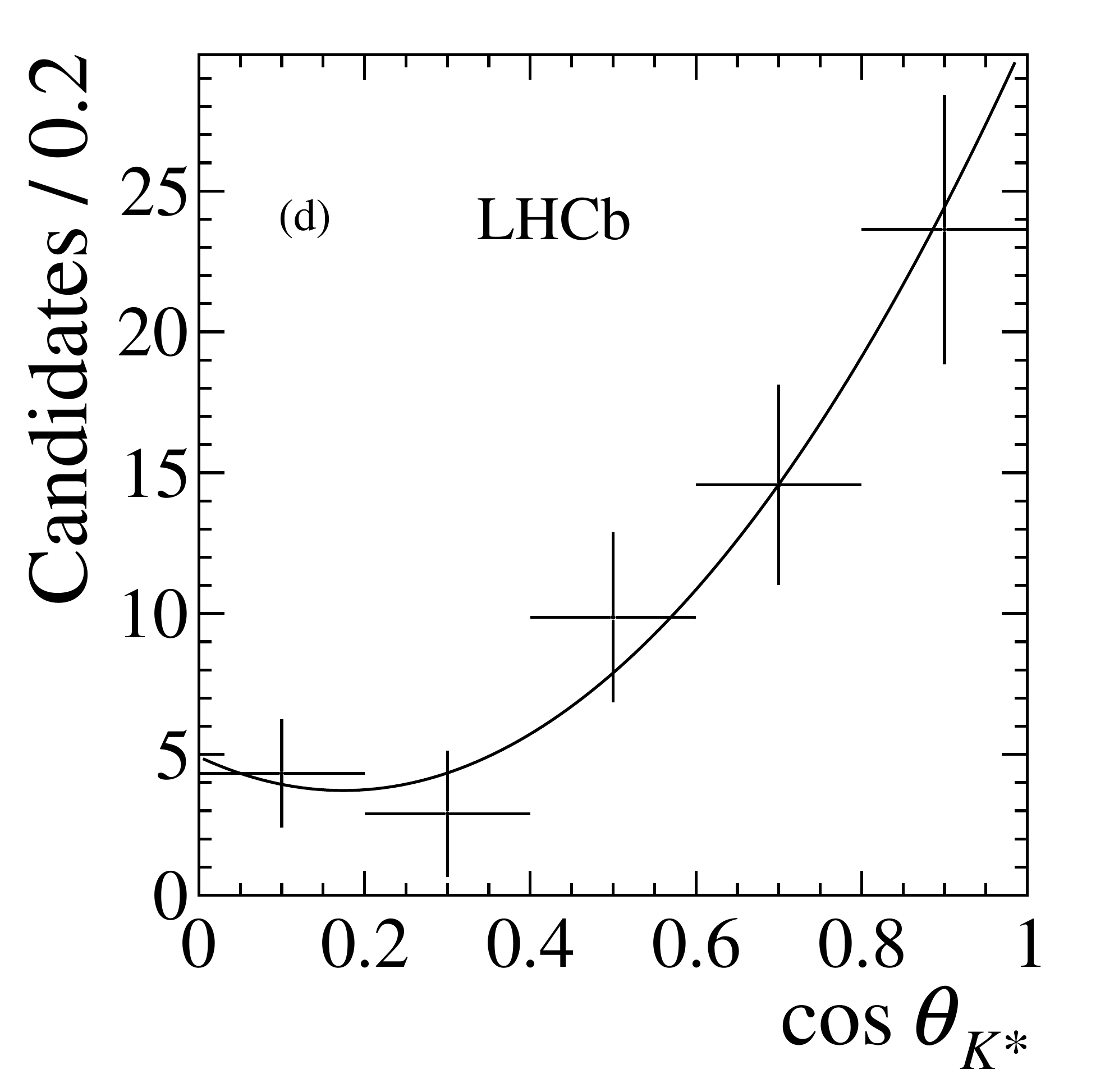}
\caption{\small
  Background-subtracted distribution of $\cos{\theta_{\Kstar}}$ for (a,b) \BdtoKstarppim and (c,d) \BstoKstarpmKmp signals from the samples with (a,c) long and (b,d) downstream candidates.
  Results of fits with second-order polynomial functions are shown as the solid lines.
}
\label{fig:Swave}
\end{figure}

The largest source of systematic uncertainty arises due to imperfect cancellation of interference effects between the P-wave $\Kstar$ signal and the nonresonant component, in which the $\KS\pipm$ system is predominantly S-wave.
Since the efficiency is not uniform as a function of the cosine of the decay angle, $\cos{\theta_{\Kstar}}$, defined as the angle between the \B and \KS candidate momenta in the rest frame of the $\KS\pipm$ system, a residual interference effect may bias the results.
The size of this uncertainty is evaluated by fitting the distribution of $\cos{\theta_{\Kstar}}$~\cite{LHCb-PAPER-2013-048}.
The distribution is reconstructed from the signal \sWeights~\cite{Pivk:2004ty} obtained from the default fit.
Only the region where $\cos{\theta_{\Kstar}}$ is positive is considered, since the efficiency variation is highly non-trivial in the negative region.
This ensures that the assigned uncertainty is conservative since any
cancellation of the interference effects between the two sides of the
distribution is neglected.
In the absence of interference, the distribution will be parabolic and pass through the origin.
The bias on the signal yield due to interference can therefore be evaluated from the constant and linear components resulting from a fit of the distribution to a second-order polynomial.
Such fits are shown for \BdtoKstarppim and \BstoKstarpmKmp signals in Fig.~\ref{fig:Swave}.
The measured yields of the \BstoKstarmpip and \BdtoKstarpmKmp signals are too small to allow this method to be used.
Therefore the same relative uncertainties are assigned to these decays as in the corresponding \Bd or \Bs decay.

Systematic uncertainties on the ratio of efficiencies arise due to limited sizes of the simulation samples used to determine the acceptance and selection efficiencies, and due to possible mismodelling of the detector response.
Two potential sources of mismodelling are the trigger and particle identification efficiencies.  
These are determined from control samples and systematic uncertainties assigned using the same procedures as described in Ref.~\cite{LHCb-PAPER-2013-042}.
The imperfect knowledge of the ratio of fragmentation fractions, $f_s/f_d = 0.259 \pm 0.015$~\cite{LHCb-PAPER-2011-018,LHCb-PAPER-2012-037,LHCb-CONF-2013-011}, is another source of uncertainty.

\begin{table}[!tb]
  \centering
  \caption{\small
    Systematic uncertainties on the relative branching fraction measurements.
    The total uncertainty is obtained by combining all sources in quadrature.
}
  \label{tab:systematics}
  \resizebox{\textwidth}{!}{
  \begin{tabular}{lcccccc}
    \hline
    Source &
    \multicolumn{2}{c}{$\frac{\Br{\BstoKstarpmKmp}}{\Br{\BdtoKstarppim}}$} & 
    \multicolumn{2}{c}{$\frac{\Br{\BdtoKstarpmKmp}}{\Br{\BdtoKstarppim}}$} & 
    \multicolumn{2}{c}{$\frac{\Br{\BstoKstarmpip}}{\Br{\BdtoKstarppim}}$} \\
    & long & downstream & long & downstream & long & downstream \\
    \hline
    Fit                     & $0.14$ & $0.07$ & $\phantom{<\,}0.010$ & $\phantom{<\,}0.005$ & $\phantom{<\,}0.05$ & $0.04$ \\
    S-wave interference     & $0.32$ & $0.14$ & $\phantom{<\,}0.001$ & $\phantom{<\,}0.002$ & $\phantom{<\,}0.04$ & $0.05$ \\
    Acceptance              & $0.01$ & $0.01$ & $<0.001$ & $<0.001$ & $<0.01$ & $0.01$ \\
    Selection               & $0.08$ & $0.05$ & $<0.001$ & $\phantom{<\,}0.001$ & $\phantom{<\,}0.01$ & $0.02$ \\
    Trigger                 & $0.03$ & $0.02$ & $<0.001$ & $\phantom{<\,}0.001$ & $<0.01$ & $0.01$ \\
    Particle identification & $0.04$ & $0.03$ & $<0.001$ & $\phantom{<\,}0.001$ & $\phantom{<\,}0.01$ & $0.01$ \\
    $f_{s}/f_{d} $          & $0.10$ & $0.08$ & ---      & ---                & $\phantom{<\,}0.01$ & $0.03$ \\
    \hline
    Total                   & $0.37$ & $0.19$ & $\phantom{<\,}0.011$ & $\phantom{<\,}0.006$ & $\phantom{<\,}0.06$ & $0.08$ \\
    \hline
  \end{tabular}
}
\end{table}

\section{Results and conclusion}
\label{sec:results}

The significance of the signal strengths is determined from
$\sqrt{-2\Delta\ln {\cal L}}$,
where $\Delta\ln {\cal L}$ is the change in the log likelihood between the default fit result and that obtained when the relevant component is fixed to zero.
This calculation is performed both with only the statistical uncertainty included, and after the likelihood function is convolved with a Gaussian function with width corresponding to the systematic uncertainty on the fitted yield.
Combining the likelihoods from long and downstream categories, the statistical significances for \BstoKstarpmKmp and \BstoKstarmpip decays are $12.5\,\sigma$ and $3.9\,\sigma$ while the corresponding values for the total significance are $7.8\,\sigma$ and $3.4\,\sigma$, respectively.
The significance of the \BdtoKstarpmKmp signal is below $2\,\sigma$.

The ratios of branching fractions of Eqs.~(\ref{eq:master-formula-Bs},\ref{eq:master-formula-Bd}) are obtained by correcting the ratios of yields by the ratios of efficiencies and, where appropriate, by the ratio of the fragmentation fractions.
The particle identification efficiencies are determined from data, using samples of kaons and pions from $\Dstarp\to\Dz\pip, \Dz\to\Km\pip$ decays reweighted according to the kinematic distributions of the bachelor tracks in $\Bds\to\Kstarpm\hadmp$ decays.
The relative efficiencies of the acceptance and all other selection requirements are determined from simulation.
The relative efficiencies are within 10\,\% of unity.

Since the signal for \BdtoKstarpmKmp decays is not significant, upper limits
at 90\,\% and 95\,\% confidence level (CL) are obtained by integrating the profile likelihood function in the region of positive branching fraction.
All results from the samples with long and downstream candidates are consistent and the combined results are 
\begin{eqnarray*}
\frac{\Br{\BstoKstarpmKmp}}{\Br{\BdtoKstarppim}} & = & 1.49\pm0.22 \stat \pm0.18 \syst \, ,\\
\frac{\Br{\BdtoKstarpmKmp}}{\Br{\BdtoKstarppim}} & = & 0.02\pm0.02 \stat \pm0.01 \syst \, ,\\
     & < & 0.05 \ (0.06) \ {\rm at} \ 90\,\% \ (95\,\%) \ {\rm CL} \, , \\
\frac{\Br{\BstoKstarmpip}}{\Br{\BdtoKstarppim}}  & = & 0.39\pm0.13 \stat \pm0.05 \syst \, .
\end{eqnarray*}
Multiplying the relative branching fractions by ${\cal B}\left(\Bd\to\Kstarp\pim\right) = \left(8.5\pm0.7\right)\times 10^{-6}$~\cite{HFAG} gives
\begin{eqnarray*}
\Br{\BstoKstarpmKmp} & = & \left( 12.7\pm1.9 \stat \pm1.9 \syst \right) \tentimes{-6} \, ,\\
\Br{\BdtoKstarpmKmp} & = & \left( 0.17\pm0.15 \stat \pm0.05 \syst \right) \tentimes{-6} \, , \\ 
                     & < & 0.4 \ (0.5) \tentimes{-6} \ {\rm at} \ 90\,\% \ (95\,\%) \ {\rm CL} \, , \\
\Br{\BstoKstarmpip}  & = & \left( 3.3\pm1.1 \stat \pm0.5 \syst \right) \tentimes{-6} \, . \\
\end{eqnarray*}

In summary, $\Bds\to\Kstarpm\hadmp$ decays have been studied using a data sample corresponding to $1.0\invfb$ of $pp$ collision data at a centre-of-mass energy of $7\tev$ collected with the LHCb detector.
The first observation of the \BstoKstarpmKmp decay and the first evidence for the \BstoKstarmpip decay are obtained, and their branching fractions measured.  
An upper limit is set on the branching fraction of the \BdtoKstarpmKmp decay.
The results are consistent with several independent theoretical predictions~\cite{Cheng:2009mu,Ali:2007ff,Su:2011eq}, and represent an important step towards searches for physics beyond the Standard Model in decays of \B mesons to charmless final states containing one pseudoscalar and one vector meson.
Dalitz plot analyses of larger samples will allow the reduction of both statistical and systematic uncertainties on these results.
The additional sensitivity to relative phases provided by such analyses will also
permit searches for sources of \CP violation beyond the Standard Model.

\section*{Acknowledgements}

\noindent We express our gratitude to our colleagues in the CERN
accelerator departments for the excellent performance of the LHC. We
thank the technical and administrative staff at the LHCb
institutes. We acknowledge support from CERN and from the national
agencies: CAPES, CNPq, FAPERJ and FINEP (Brazil); NSFC (China);
CNRS/IN2P3 (France); BMBF, DFG, HGF and MPG (Germany); SFI (Ireland); INFN (Italy); 
FOM and NWO (The Netherlands); MNiSW and NCN (Poland); MEN/IFA (Romania); 
MinES and FANO (Russia); MinECo (Spain); SNSF and SER (Switzerland); 
NASU (Ukraine); STFC (United Kingdom); NSF (USA).
The Tier1 computing centres are supported by IN2P3 (France), KIT and BMBF 
(Germany), INFN (Italy), NWO and SURF (The Netherlands), PIC (Spain), GridPP 
(United Kingdom).
We are indebted to the communities behind the multiple open 
source software packages on which we depend. We are also thankful for the 
computing resources and the access to software R\&D tools provided by Yandex LLC (Russia).
Individual groups or members have received support from 
EPLANET, Marie Sk\l{}odowska-Curie Actions and ERC (European Union), 
Conseil g\'{e}n\'{e}ral de Haute-Savoie, Labex ENIGMASS and OCEVU, 
R\'{e}gion Auvergne (France), RFBR (Russia), XuntaGal and GENCAT (Spain), Royal Society and Royal
Commission for the Exhibition of 1851 (United Kingdom).

\addcontentsline{toc}{section}{References}
\setboolean{inbibliography}{true}
\bibliographystyle{LHCb}
\bibliography{main,LHCb-PAPER,LHCb-CONF,LHCb-DP}

\ifx\mcitethebibliography\mciteundefinedmacro
\PackageError{LHCb.bst}{mciteplus.sty has not been loaded}
{This bibstyle requires the use of the mciteplus package.}\fi
\providecommand{\href}[2]{#2}
\begin{mcitethebibliography}{10}
\mciteSetBstSublistMode{n}
\mciteSetBstMaxWidthForm{subitem}{\alph{mcitesubitemcount})}
\mciteSetBstSublistLabelBeginEnd{\mcitemaxwidthsubitemform\space}
{\relax}{\relax}

\bibitem{Cabibbo:1963yz}
N.~Cabibbo, \ifthenelse{\boolean{articletitles}}{{\it {Unitary symmetry and
  leptonic decays}},
  }{}\href{http://dx.doi.org/10.1103/PhysRevLett.10.531}{Phys.\ Rev.\ Lett.\
  {\bf 10} (1963) 531}\relax
\mciteBstWouldAddEndPuncttrue
\mciteSetBstMidEndSepPunct{\mcitedefaultmidpunct}
{\mcitedefaultendpunct}{\mcitedefaultseppunct}\relax
\EndOfBibitem
\bibitem{Kobayashi:1973fv}
M.~Kobayashi and T.~Maskawa, \ifthenelse{\boolean{articletitles}}{{\it {\CP
  violation in the renormalizable theory of weak interaction}},
  }{}\href{http://dx.doi.org/10.1143/PTP.49.652}{Prog.\ Theor.\ Phys.\  {\bf
  49} (1973) 652}\relax
\mciteBstWouldAddEndPuncttrue
\mciteSetBstMidEndSepPunct{\mcitedefaultmidpunct}
{\mcitedefaultendpunct}{\mcitedefaultseppunct}\relax
\EndOfBibitem
\bibitem{Riotto:1999yt}
A.~Riotto and M.~Trodden, \ifthenelse{\boolean{articletitles}}{{\it {Recent
  progress in baryogenesis}},
  }{}\href{http://dx.doi.org/10.1146/annurev.nucl.49.1.35}{Ann.\ Rev.\ Nucl.\
  Part.\ Sci.\  {\bf 49} (1999) 35},
  \href{http://arxiv.org/abs/hep-ph/9901362}{{\tt arXiv:hep-ph/9901362}}\relax
\mciteBstWouldAddEndPuncttrue
\mciteSetBstMidEndSepPunct{\mcitedefaultmidpunct}
{\mcitedefaultendpunct}{\mcitedefaultseppunct}\relax
\EndOfBibitem
\bibitem{Antonelli:2009ws}
M.~Antonelli {\em et~al.}, \ifthenelse{\boolean{articletitles}}{{\it {Flavor
  physics in the quark sector}},
  }{}\href{http://dx.doi.org/10.1016/j.physrep.2010.05.003}{Phys.\ Rept.\  {\bf
  494} (2010) 197}, \href{http://arxiv.org/abs/0907.5386}{{\tt
  arXiv:0907.5386}}\relax
\mciteBstWouldAddEndPuncttrue
\mciteSetBstMidEndSepPunct{\mcitedefaultmidpunct}
{\mcitedefaultendpunct}{\mcitedefaultseppunct}\relax
\EndOfBibitem
\bibitem{Lees:2012kx}
\babar collaboration, J.~P. Lees {\em et~al.},
  \ifthenelse{\boolean{articletitles}}{{\it {Measurement of \CP asymmetries and
  branching fractions in charmless two-body $B$-meson decays to pions and
  kaons}}, }{}\href{http://dx.doi.org/10.1103/PhysRevD.87.052009}{Phys.\ Rev.\
  {\bf D87} (2013) 052009}, \href{http://arxiv.org/abs/1206.3525}{{\tt
  arXiv:1206.3525}}\relax
\mciteBstWouldAddEndPuncttrue
\mciteSetBstMidEndSepPunct{\mcitedefaultmidpunct}
{\mcitedefaultendpunct}{\mcitedefaultseppunct}\relax
\EndOfBibitem
\bibitem{Duh:2012ie}
Belle collaboration, Y.-T. Duh {\em et~al.},
  \ifthenelse{\boolean{articletitles}}{{\it {Measurements of branching
  fractions and direct \CP asymmetries for $B \to K\pi$, $B\to \pi\pi$ and
  $B\to KK$ decays}},
  }{}\href{http://dx.doi.org/10.1103/PhysRevD.87.031103}{Phys.\ Rev.\  {\bf
  D87} (2013) 031103}, \href{http://arxiv.org/abs/1210.1348}{{\tt
  arXiv:1210.1348}}\relax
\mciteBstWouldAddEndPuncttrue
\mciteSetBstMidEndSepPunct{\mcitedefaultmidpunct}
{\mcitedefaultendpunct}{\mcitedefaultseppunct}\relax
\EndOfBibitem
\bibitem{LHCb-PAPER-2013-018}
LHCb collaboration, R.~Aaij {\em et~al.},
  \ifthenelse{\boolean{articletitles}}{{\it {First observation of $CP$
  violation in the decays of $B_s^0$ mesons}},
  }{}\href{http://dx.doi.org/10.1103/PhysRevLett.110.221601}{Phys.\ Rev.\
  Lett.\  {\bf 110} (2013) 221601}, \href{http://arxiv.org/abs/1304.6173}{{\tt
  arXiv:1304.6173}}\relax
\mciteBstWouldAddEndPuncttrue
\mciteSetBstMidEndSepPunct{\mcitedefaultmidpunct}
{\mcitedefaultendpunct}{\mcitedefaultseppunct}\relax
\EndOfBibitem
\bibitem{Aaltonen:2014vra}
CDF collaboration, T.~A. Aaltonen {\em et~al.},
  \ifthenelse{\boolean{articletitles}}{{\it {Measurements of direct
  \CP-violating asymmetries in charmless decays of bottom baryons}},
  }{}\href{http://arxiv.org/abs/1403.5586}{{\tt arXiv:1403.5586}}\relax
\mciteBstWouldAddEndPuncttrue
\mciteSetBstMidEndSepPunct{\mcitedefaultmidpunct}
{\mcitedefaultendpunct}{\mcitedefaultseppunct}\relax
\EndOfBibitem
\bibitem{LHCb-PAPER-2013-027}
LHCb collaboration, R.~Aaij {\em et~al.},
  \ifthenelse{\boolean{articletitles}}{{\it {Measurement of $CP$ violation in
  the phase space of $B^\pm \to K^\pm\pi^+\pi^-$ and $B^\pm \to K^\pm K^+K^-$
  decays}}, }{}\href{http://dx.doi.org/10.1103/PhysRevLett.111.101801}{Phys.\
  Rev.\ Lett.\  {\bf 111} (2013) 101801},
  \href{http://arxiv.org/abs/1306.1246}{{\tt arXiv:1306.1246}}\relax
\mciteBstWouldAddEndPuncttrue
\mciteSetBstMidEndSepPunct{\mcitedefaultmidpunct}
{\mcitedefaultendpunct}{\mcitedefaultseppunct}\relax
\EndOfBibitem
\bibitem{LHCb-PAPER-2013-051}
LHCb collaboration, R.~Aaij {\em et~al.},
  \ifthenelse{\boolean{articletitles}}{{\it {Measurement of $CP$ violation in
  the phase space of $B^\pm \to K^+K^-\pi^\pm$ and $B^\pm\to \pi^+\pi^-\pi^\pm$
  decays}}, }{}\href{http://dx.doi.org/10.1103/PhysRevLett.112.011801}{Phys.\
  Rev.\ Lett.\  {\bf 112} (2014) 011801},
  \href{http://arxiv.org/abs/1310.4740}{{\tt arXiv:1310.4740}}\relax
\mciteBstWouldAddEndPuncttrue
\mciteSetBstMidEndSepPunct{\mcitedefaultmidpunct}
{\mcitedefaultendpunct}{\mcitedefaultseppunct}\relax
\EndOfBibitem
\bibitem{LHCb-PAPER-2014-044}
LHCb collaboration, R.~Aaij {\em et~al.},
  \ifthenelse{\boolean{articletitles}}{{\it {Measurement of $CP$ violation in
  the three-body phase space of charmless $B^\pm$ decays}},
  }{}\href{http://arxiv.org/abs/1408.5373}{{\tt arXiv:1408.5373}}, {submitted
  to Phys. Rev. D}\relax
\mciteBstWouldAddEndPuncttrue
\mciteSetBstMidEndSepPunct{\mcitedefaultmidpunct}
{\mcitedefaultendpunct}{\mcitedefaultseppunct}\relax
\EndOfBibitem
\bibitem{Dalitz:1953cp}
R.~H. Dalitz, \ifthenelse{\boolean{articletitles}}{{\it {On the analysis of
  tau-meson data and the nature of the tau-meson}},
  }{}\href{http://dx.doi.org/10.1080/14786441008520365}{Phil.\ Mag.\  {\bf 44}
  (1953) 1068}\relax
\mciteBstWouldAddEndPuncttrue
\mciteSetBstMidEndSepPunct{\mcitedefaultmidpunct}
{\mcitedefaultendpunct}{\mcitedefaultseppunct}\relax
\EndOfBibitem
\bibitem{Ciuchini:2006kv}
M.~Ciuchini, M.~Pierini, and L.~Silvestrini,
  \ifthenelse{\boolean{articletitles}}{{\it {New bounds on the
  Cabibbo-Kobayashi-Maskawa matrix from $B \to K \pi \pi$ Dalitz plot
  analyses}}, }{}\href{http://dx.doi.org/10.1103/PhysRevD.74.051301}{Phys.\
  Rev.\  {\bf D74} (2006) 051301},
  \href{http://arxiv.org/abs/hep-ph/0601233}{{\tt arXiv:hep-ph/0601233}}\relax
\mciteBstWouldAddEndPuncttrue
\mciteSetBstMidEndSepPunct{\mcitedefaultmidpunct}
{\mcitedefaultendpunct}{\mcitedefaultseppunct}\relax
\EndOfBibitem
\bibitem{Ciuchini:2006st}
M.~Ciuchini, M.~Pierini, and L.~Silvestrini,
  \ifthenelse{\boolean{articletitles}}{{\it {Hunting the CKM weak phase with
  time-integrated Dalitz analyses of $\Bs \to K \pi \pi$ decays}},
  }{}\href{http://dx.doi.org/10.1016/j.physletb.2006.12.043}{Phys.\ Lett.\
  {\bf B645} (2007) 201}, \href{http://arxiv.org/abs/hep-ph/0602207}{{\tt
  arXiv:hep-ph/0602207}}\relax
\mciteBstWouldAddEndPuncttrue
\mciteSetBstMidEndSepPunct{\mcitedefaultmidpunct}
{\mcitedefaultendpunct}{\mcitedefaultseppunct}\relax
\EndOfBibitem
\bibitem{Gronau:2006qn}
M.~Gronau, D.~Pirjol, A.~Soni, and J.~Zupan,
  \ifthenelse{\boolean{articletitles}}{{\it {Improved method for CKM
  constraints in charmless three-body \B and \Bs decays}},
  }{}\href{http://dx.doi.org/10.1103/PhysRevD.75.014002}{Phys.\ Rev.\  {\bf
  D75} (2007) 014002}, \href{http://arxiv.org/abs/hep-ph/0608243}{{\tt
  arXiv:hep-ph/0608243}}\relax
\mciteBstWouldAddEndPuncttrue
\mciteSetBstMidEndSepPunct{\mcitedefaultmidpunct}
{\mcitedefaultendpunct}{\mcitedefaultseppunct}\relax
\EndOfBibitem
\bibitem{Gronau:2007vr}
M.~Gronau, D.~Pirjol, A.~Soni, and J.~Zupan,
  \ifthenelse{\boolean{articletitles}}{{\it {Constraint on $\bar{\rho},
  \bar{\eta}$ from $B \to K^* \pi$}},
  }{}\href{http://dx.doi.org/10.1103/PhysRevD.77.057504}{Phys.\ Rev.\  {\bf
  D77} (2008) 057504}, \href{http://arxiv.org/abs/0712.3751}{{\tt
  arXiv:0712.3751}}\relax
\mciteBstWouldAddEndPuncttrue
\mciteSetBstMidEndSepPunct{\mcitedefaultmidpunct}
{\mcitedefaultendpunct}{\mcitedefaultseppunct}\relax
\EndOfBibitem
\bibitem{Bediaga:2006jk}
I.~Bediaga, G.~Guerrer, and J.~M. de~Miranda,
  \ifthenelse{\boolean{articletitles}}{{\it {Extracting the quark mixing phase
  $\gamma$ from $\Bpm \to \Kpm \pip \pim$, $\Bz \to \KS \pip \pim$, and $\Bzb
  \to \KS \pip \pim$}},
  }{}\href{http://dx.doi.org/10.1103/PhysRevD.76.073011}{Phys.\ Rev.\  {\bf
  D76} (2007) 073011}, \href{http://arxiv.org/abs/hep-ph/0608268}{{\tt
  arXiv:hep-ph/0608268}}\relax
\mciteBstWouldAddEndPuncttrue
\mciteSetBstMidEndSepPunct{\mcitedefaultmidpunct}
{\mcitedefaultendpunct}{\mcitedefaultseppunct}\relax
\EndOfBibitem
\bibitem{Gronau:2010dd}
M.~Gronau, D.~Pirjol, and J.~Zupan, \ifthenelse{\boolean{articletitles}}{{\it
  {\CP asymmetries in $B \rightarrow K\pi, K^*\pi, \rho K $ decays}},
  }{}\href{http://dx.doi.org/10.1103/PhysRevD.81.094011}{Phys.\ Rev.\  {\bf
  D81} (2010) 094011}, \href{http://arxiv.org/abs/1001.0702}{{\tt
  arXiv:1001.0702}}\relax
\mciteBstWouldAddEndPuncttrue
\mciteSetBstMidEndSepPunct{\mcitedefaultmidpunct}
{\mcitedefaultendpunct}{\mcitedefaultseppunct}\relax
\EndOfBibitem
\bibitem{Garmash:2005rv}
Belle collaboration, A.~Garmash {\em et~al.},
  \ifthenelse{\boolean{articletitles}}{{\it {Evidence for large direct \CP
  violation in $\Bpm \to \rho(770)^0\Kpm$ from analysis of the three-body
  charmless $\Bpm \to \Kpm \pipm\pimp$ decays}},
  }{}\href{http://dx.doi.org/10.1103/PhysRevLett.96.251803}{Phys.\ Rev.\ Lett.\
   {\bf 96} (2006) 251803}, \href{http://arxiv.org/abs/hep-ex/0512066}{{\tt
  arXiv:hep-ex/0512066}}\relax
\mciteBstWouldAddEndPuncttrue
\mciteSetBstMidEndSepPunct{\mcitedefaultmidpunct}
{\mcitedefaultendpunct}{\mcitedefaultseppunct}\relax
\EndOfBibitem
\bibitem{Aubert:2008bj}
\babar collaboration, B.~Aubert {\em et~al.},
  \ifthenelse{\boolean{articletitles}}{{\it {Evidence for direct \CP violation
  from Dalitz-plot analysis of $B^\pm \to K^\pm \pi^\mp \pi^\pm$}},
  }{}\href{http://dx.doi.org/10.1103/PhysRevD.78.012004}{Phys.\ Rev.\  {\bf
  D78} (2008) 012004}, \href{http://arxiv.org/abs/0803.4451}{{\tt
  arXiv:0803.4451}}\relax
\mciteBstWouldAddEndPuncttrue
\mciteSetBstMidEndSepPunct{\mcitedefaultmidpunct}
{\mcitedefaultendpunct}{\mcitedefaultseppunct}\relax
\EndOfBibitem
\bibitem{:2008wwa}
Belle collaboration, J.~Dalseno {\em et~al.},
  \ifthenelse{\boolean{articletitles}}{{\it {Time-dependent Dalitz plot
  measurement of \CP parameters in $B^0 \to \KS \pi^+ \pi^-$ decays}},
  }{}\href{http://dx.doi.org/10.1103/PhysRevD.79.072004}{Phys.\ Rev.\  {\bf
  D79} (2009) 072004}, \href{http://arxiv.org/abs/0811.3665}{{\tt
  arXiv:0811.3665}}\relax
\mciteBstWouldAddEndPuncttrue
\mciteSetBstMidEndSepPunct{\mcitedefaultmidpunct}
{\mcitedefaultendpunct}{\mcitedefaultseppunct}\relax
\EndOfBibitem
\bibitem{Aubert:2009me}
\babar collaboration, B.~Aubert {\em et~al.},
  \ifthenelse{\boolean{articletitles}}{{\it {Time-dependent amplitude analysis
  of $B^0 \to \KS \pi^+ \pi^-$}},
  }{}\href{http://dx.doi.org/10.1103/PhysRevD.80.112001}{Phys.\ Rev.\  {\bf
  D80} (2009) 112001}, \href{http://arxiv.org/abs/0905.3615}{{\tt
  arXiv:0905.3615}}\relax
\mciteBstWouldAddEndPuncttrue
\mciteSetBstMidEndSepPunct{\mcitedefaultmidpunct}
{\mcitedefaultendpunct}{\mcitedefaultseppunct}\relax
\EndOfBibitem
\bibitem{BABAR:2011ae}
\babar collaboration, J.~P. Lees {\em et~al.},
  \ifthenelse{\boolean{articletitles}}{{\it {Amplitude analysis of $B^0\to K^+
  \pi^- \pi^0$ and evidence of direct \CP violation in $B\to K^* \pi$ decays}},
  }{}\href{http://dx.doi.org/10.1103/PhysRevD.83.112010}{Phys.\ Rev.\  {\bf
  D83} (2011) 112010}, \href{http://arxiv.org/abs/1105.0125}{{\tt
  arXiv:1105.0125}}\relax
\mciteBstWouldAddEndPuncttrue
\mciteSetBstMidEndSepPunct{\mcitedefaultmidpunct}
{\mcitedefaultendpunct}{\mcitedefaultseppunct}\relax
\EndOfBibitem
\bibitem{Aubert:2006wu}
\babar collaboration, B.~Aubert {\em et~al.},
  \ifthenelse{\boolean{articletitles}}{{\it {Search for the decay of a \Bd or
  \Bdb meson to $\Kstarzb\Kz$ or $\Kstarz\Kzb$}},
  }{}\href{http://dx.doi.org/10.1103/PhysRevD.74.072008}{Phys.\ Rev.\  {\bf
  D74} (2006) 072008}, \href{http://arxiv.org/abs/hep-ex/0606050}{{\tt
  arXiv:hep-ex/0606050}}\relax
\mciteBstWouldAddEndPuncttrue
\mciteSetBstMidEndSepPunct{\mcitedefaultmidpunct}
{\mcitedefaultendpunct}{\mcitedefaultseppunct}\relax
\EndOfBibitem
\bibitem{Aubert:2007ua}
\babar collaboration, B.~Aubert {\em et~al.},
  \ifthenelse{\boolean{articletitles}}{{\it {Search for the decay $\Bp \to
  \Kstarb(892)^0 \Kp$}},
  }{}\href{http://dx.doi.org/10.1103/PhysRevD.76.071103}{Phys.\ Rev.\  {\bf
  D76} (2007) 071103}, \href{http://arxiv.org/abs/0706.1059}{{\tt
  arXiv:0706.1059}}\relax
\mciteBstWouldAddEndPuncttrue
\mciteSetBstMidEndSepPunct{\mcitedefaultmidpunct}
{\mcitedefaultendpunct}{\mcitedefaultseppunct}\relax
\EndOfBibitem
\bibitem{Aubert:2007xb}
\babar collaboration, B.~Aubert {\em et~al.},
  \ifthenelse{\boolean{articletitles}}{{\it {Observation of the decay $B^{+}
  \to K^{+} K^{-} \pi^{+}$}},
  }{}\href{http://dx.doi.org/10.1103/PhysRevLett.99.221801}{Phys.\ Rev.\ Lett.\
   {\bf 99} (2007) 221801}, \href{http://arxiv.org/abs/0708.0376}{{\tt
  arXiv:0708.0376}}\relax
\mciteBstWouldAddEndPuncttrue
\mciteSetBstMidEndSepPunct{\mcitedefaultmidpunct}
{\mcitedefaultendpunct}{\mcitedefaultseppunct}\relax
\EndOfBibitem
\bibitem{Aubert:2008aw}
\babar collaboration, B.~Aubert {\em et~al.},
  \ifthenelse{\boolean{articletitles}}{{\it {Search for the decay $\Bp \to
  \KS\KS\pip$}}, }{}\href{http://dx.doi.org/10.1103/PhysRevD.79.051101}{Phys.\
  Rev.\  {\bf D79} (2009) 051101}, \href{http://arxiv.org/abs/0811.1979}{{\tt
  arXiv:0811.1979}}\relax
\mciteBstWouldAddEndPuncttrue
\mciteSetBstMidEndSepPunct{\mcitedefaultmidpunct}
{\mcitedefaultendpunct}{\mcitedefaultseppunct}\relax
\EndOfBibitem
\bibitem{delAmoSanchez:2010ur}
\babar collaboration, P.~del Amo~Sanchez {\em et~al.},
  \ifthenelse{\boolean{articletitles}}{{\it {Observation of the rare decay $\Bd
  \to \KS\Kpm\pimp$}},
  }{}\href{http://dx.doi.org/10.1103/PhysRevD.82.031101}{Phys.\ Rev.\  {\bf
  D82} (2010) 031101}, \href{http://arxiv.org/abs/1003.0640}{{\tt
  arXiv:1003.0640}}\relax
\mciteBstWouldAddEndPuncttrue
\mciteSetBstMidEndSepPunct{\mcitedefaultmidpunct}
{\mcitedefaultendpunct}{\mcitedefaultseppunct}\relax
\EndOfBibitem
\bibitem{Gaur:2013uou}
Belle collaboration, V.~Gaur {\em et~al.},
  \ifthenelse{\boolean{articletitles}}{{\it {Evidence for the decay $B^0 \to
  K^+K^-\pi^0$}}, }{}\href{http://dx.doi.org/10.1103/PhysRevD.87.091101}{Phys.\
  Rev.\  {\bf D87} (2013) 091101}, \href{http://arxiv.org/abs/1304.5312}{{\tt
  arXiv:1304.5312}}\relax
\mciteBstWouldAddEndPuncttrue
\mciteSetBstMidEndSepPunct{\mcitedefaultmidpunct}
{\mcitedefaultendpunct}{\mcitedefaultseppunct}\relax
\EndOfBibitem
\bibitem{LHCb-PAPER-2013-042}
LHCb collaboration, R.~Aaij {\em et~al.},
  \ifthenelse{\boolean{articletitles}}{{\it {Study of $B_{(s)}^0 \to K^0_S h^+
  h^{\prime -}$ decays with first observation of $B_s^0 \to K^0_S K^\pm
  \pi^\mp$ and $B_s^0 \to K^0_S \pi^+\pi^-$}},
  }{}\href{http://dx.doi.org/10.1007/JHEP10(2013)143}{JHEP {\bf 10} (2013)
  143}, \href{http://arxiv.org/abs/1307.7648}{{\tt arXiv:1307.7648}}\relax
\mciteBstWouldAddEndPuncttrue
\mciteSetBstMidEndSepPunct{\mcitedefaultmidpunct}
{\mcitedefaultendpunct}{\mcitedefaultseppunct}\relax
\EndOfBibitem
\bibitem{HFAG}
Heavy Flavor Averaging Group, Y.~Amhis {\em et~al.},
  \ifthenelse{\boolean{articletitles}}{{\it {Averages of $b$-hadron,
  $c$-hadron, and $\tau$-lepton properties as of early 2012}},
  }{}\href{http://arxiv.org/abs/1207.1158}{{\tt arXiv:1207.1158}}, {updated
  results and plots available at
  \href{http://www.slac.stanford.edu/xorg/hfag/}{{\tt
  http://www.slac.stanford.edu/xorg/hfag/}}}\relax
\mciteBstWouldAddEndPuncttrue
\mciteSetBstMidEndSepPunct{\mcitedefaultmidpunct}
{\mcitedefaultendpunct}{\mcitedefaultseppunct}\relax
\EndOfBibitem
\bibitem{LHCb-PAPER-2011-018}
LHCb collaboration, R.~Aaij {\em et~al.},
  \ifthenelse{\boolean{articletitles}}{{\it {Measurement of $b$ hadron
  production fractions in 7 TeV $pp$ collisions}},
  }{}\href{http://dx.doi.org/10.1103/PhysRevD.85.032008}{Phys.\ Rev.\  {\bf
  D85} (2012) 032008}, \href{http://arxiv.org/abs/1111.2357}{{\tt
  arXiv:1111.2357}}\relax
\mciteBstWouldAddEndPuncttrue
\mciteSetBstMidEndSepPunct{\mcitedefaultmidpunct}
{\mcitedefaultendpunct}{\mcitedefaultseppunct}\relax
\EndOfBibitem
\bibitem{LHCb-PAPER-2012-037}
LHCb collaboration, R.~Aaij {\em et~al.},
  \ifthenelse{\boolean{articletitles}}{{\it {Measurement of the fragmentation
  fraction ratio $f_s/f_d$ and its dependence on $B$ meson kinematics}},
  }{}\href{http://dx.doi.org/10.1007/JHEP04(2013)001}{JHEP {\bf 04} (2013)
  001}, \href{http://arxiv.org/abs/1301.5286}{{\tt arXiv:1301.5286}}\relax
\mciteBstWouldAddEndPuncttrue
\mciteSetBstMidEndSepPunct{\mcitedefaultmidpunct}
{\mcitedefaultendpunct}{\mcitedefaultseppunct}\relax
\EndOfBibitem
\bibitem{LHCb-CONF-2013-011}
{LHCb collaboration}, \ifthenelse{\boolean{articletitles}}{{\it {Updated
  average $f_{s}/f_{d}$ $b$-hadron production fraction ratio for $7 \tev$ $pp$
  collisions}}, }{}
  \href{http://cdsweb.cern.ch/search?p={LHCb-CONF-2013-011}&f=reportnumber&action_search=Search&c=LHCb+Reports&c=LHCb+Conference+Proceedings&c=LHCb+Conference+Contributions&c=LHCb+Notes&c=LHCb+Theses&c=LHCb+Papers}
  {{LHCb-CONF-2013-011}}\relax
\mciteBstWouldAddEndPuncttrue
\mciteSetBstMidEndSepPunct{\mcitedefaultmidpunct}
{\mcitedefaultendpunct}{\mcitedefaultseppunct}\relax
\EndOfBibitem
\bibitem{Alves:2008zz}
LHCb collaboration, A.~A. Alves~Jr. {\em et~al.},
  \ifthenelse{\boolean{articletitles}}{{\it {The \lhcb detector at the LHC}},
  }{}\href{http://dx.doi.org/10.1088/1748-0221/3/08/S08005}{JINST {\bf 3}
  (2008) S08005}\relax
\mciteBstWouldAddEndPuncttrue
\mciteSetBstMidEndSepPunct{\mcitedefaultmidpunct}
{\mcitedefaultendpunct}{\mcitedefaultseppunct}\relax
\EndOfBibitem
\bibitem{LHCb-DP-2014-001}
R.~Aaij {\em et~al.}, \ifthenelse{\boolean{articletitles}}{{\it {Performance of
  the LHCb Vertex Locator}},
  }{}\href{http://dx.doi.org/10.1088/1748-0221/9/09/P09007}{JINST {\bf 9}
  (2014) P09007}, \href{http://arxiv.org/abs/1405.7808}{{\tt
  arXiv:1405.7808}}\relax
\mciteBstWouldAddEndPuncttrue
\mciteSetBstMidEndSepPunct{\mcitedefaultmidpunct}
{\mcitedefaultendpunct}{\mcitedefaultseppunct}\relax
\EndOfBibitem
\bibitem{LHCb-DP-2013-003}
R.~Arink {\em et~al.}, \ifthenelse{\boolean{articletitles}}{{\it {Performance
  of the LHCb Outer Tracker}},
  }{}\href{http://dx.doi.org/10.1088/1748-0221/9/01/P01002}{JINST {\bf 9}
  (2014) P01002}, \href{http://arxiv.org/abs/1311.3893}{{\tt
  arXiv:1311.3893}}\relax
\mciteBstWouldAddEndPuncttrue
\mciteSetBstMidEndSepPunct{\mcitedefaultmidpunct}
{\mcitedefaultendpunct}{\mcitedefaultseppunct}\relax
\EndOfBibitem
\bibitem{LHCb-DP-2012-003}
M.~Adinolfi {\em et~al.}, \ifthenelse{\boolean{articletitles}}{{\it
  {Performance of the \lhcb RICH detector at the LHC}},
  }{}\href{http://dx.doi.org/10.1140/epjc/s10052-013-2431-9}{Eur.\ Phys.\ J.\
  {\bf C73} (2013) 2431}, \href{http://arxiv.org/abs/1211.6759}{{\tt
  arXiv:1211.6759}}\relax
\mciteBstWouldAddEndPuncttrue
\mciteSetBstMidEndSepPunct{\mcitedefaultmidpunct}
{\mcitedefaultendpunct}{\mcitedefaultseppunct}\relax
\EndOfBibitem
\bibitem{LHCb-DP-2012-002}
A.~A. Alves~Jr. {\em et~al.}, \ifthenelse{\boolean{articletitles}}{{\it
  {Performance of the LHCb muon system}},
  }{}\href{http://dx.doi.org/10.1088/1748-0221/8/02/P02022}{JINST {\bf 8}
  (2013) P02022}, \href{http://arxiv.org/abs/1211.1346}{{\tt
  arXiv:1211.1346}}\relax
\mciteBstWouldAddEndPuncttrue
\mciteSetBstMidEndSepPunct{\mcitedefaultmidpunct}
{\mcitedefaultendpunct}{\mcitedefaultseppunct}\relax
\EndOfBibitem
\bibitem{LHCb-DP-2012-004}
R.~Aaij {\em et~al.}, \ifthenelse{\boolean{articletitles}}{{\it {The \lhcb
  trigger and its performance in 2011}},
  }{}\href{http://dx.doi.org/10.1088/1748-0221/8/04/P04022}{JINST {\bf 8}
  (2013) P04022}, \href{http://arxiv.org/abs/1211.3055}{{\tt
  arXiv:1211.3055}}\relax
\mciteBstWouldAddEndPuncttrue
\mciteSetBstMidEndSepPunct{\mcitedefaultmidpunct}
{\mcitedefaultendpunct}{\mcitedefaultseppunct}\relax
\EndOfBibitem
\bibitem{BBDT}
V.~V. Gligorov and M.~Williams, \ifthenelse{\boolean{articletitles}}{{\it
  {Efficient, reliable and fast high-level triggering using a bonsai boosted
  decision tree}},
  }{}\href{http://dx.doi.org/10.1088/1748-0221/8/02/P02013}{JINST {\bf 8}
  (2013) P02013}, \href{http://arxiv.org/abs/1210.6861}{{\tt
  arXiv:1210.6861}}\relax
\mciteBstWouldAddEndPuncttrue
\mciteSetBstMidEndSepPunct{\mcitedefaultmidpunct}
{\mcitedefaultendpunct}{\mcitedefaultseppunct}\relax
\EndOfBibitem
\bibitem{Sjostrand:2006za}
T.~Sj\"{o}strand, S.~Mrenna, and P.~Skands,
  \ifthenelse{\boolean{articletitles}}{{\it {PYTHIA 6.4 physics and manual}},
  }{}\href{http://dx.doi.org/10.1088/1126-6708/2006/05/026}{JHEP {\bf 05}
  (2006) 026}, \href{http://arxiv.org/abs/hep-ph/0603175}{{\tt
  arXiv:hep-ph/0603175}}\relax
\mciteBstWouldAddEndPuncttrue
\mciteSetBstMidEndSepPunct{\mcitedefaultmidpunct}
{\mcitedefaultendpunct}{\mcitedefaultseppunct}\relax
\EndOfBibitem
\bibitem{LHCb-PROC-2010-056}
I.~Belyaev {\em et~al.}, \ifthenelse{\boolean{articletitles}}{{\it {Handling of
  the generation of primary events in \gauss, the \lhcb simulation framework}},
  }{}\href{http://dx.doi.org/10.1109/NSSMIC.2010.5873949}{Nuclear Science
  Symposium Conference Record (NSS/MIC) {\bf IEEE} (2010) 1155}\relax
\mciteBstWouldAddEndPuncttrue
\mciteSetBstMidEndSepPunct{\mcitedefaultmidpunct}
{\mcitedefaultendpunct}{\mcitedefaultseppunct}\relax
\EndOfBibitem
\bibitem{Lange:2001uf}
D.~J. Lange, \ifthenelse{\boolean{articletitles}}{{\it {The EvtGen particle
  decay simulation package}},
  }{}\href{http://dx.doi.org/10.1016/S0168-9002(01)00089-4}{Nucl.\ Instrum.\
  Meth.\  {\bf A462} (2001) 152}\relax
\mciteBstWouldAddEndPuncttrue
\mciteSetBstMidEndSepPunct{\mcitedefaultmidpunct}
{\mcitedefaultendpunct}{\mcitedefaultseppunct}\relax
\EndOfBibitem
\bibitem{Golonka:2005pn}
P.~Golonka and Z.~Was, \ifthenelse{\boolean{articletitles}}{{\it {PHOTOS Monte
  Carlo: a precision tool for QED corrections in $Z$ and $W$ decays}},
  }{}\href{http://dx.doi.org/10.1140/epjc/s2005-02396-4}{Eur.\ Phys.\ J.\  {\bf
  C45} (2006) 97}, \href{http://arxiv.org/abs/hep-ph/0506026}{{\tt
  arXiv:hep-ph/0506026}}\relax
\mciteBstWouldAddEndPuncttrue
\mciteSetBstMidEndSepPunct{\mcitedefaultmidpunct}
{\mcitedefaultendpunct}{\mcitedefaultseppunct}\relax
\EndOfBibitem
\bibitem{Allison:2006ve}
Geant4 collaboration, J.~Allison {\em et~al.},
  \ifthenelse{\boolean{articletitles}}{{\it {Geant4 developments and
  applications}}, }{}\href{http://dx.doi.org/10.1109/TNS.2006.869826}{IEEE
  Trans.\ Nucl.\ Sci.\  {\bf 53} (2006) 270}\relax
\mciteBstWouldAddEndPuncttrue
\mciteSetBstMidEndSepPunct{\mcitedefaultmidpunct}
{\mcitedefaultendpunct}{\mcitedefaultseppunct}\relax
\EndOfBibitem
\bibitem{Agostinelli:2002hh}
Geant4 collaboration, S.~Agostinelli {\em et~al.},
  \ifthenelse{\boolean{articletitles}}{{\it {Geant4: a simulation toolkit}},
  }{}\href{http://dx.doi.org/10.1016/S0168-9002(03)01368-8}{Nucl.\ Instrum.\
  Meth.\  {\bf A506} (2003) 250}\relax
\mciteBstWouldAddEndPuncttrue
\mciteSetBstMidEndSepPunct{\mcitedefaultmidpunct}
{\mcitedefaultendpunct}{\mcitedefaultseppunct}\relax
\EndOfBibitem
\bibitem{LHCb-PROC-2011-006}
M.~Clemencic {\em et~al.}, \ifthenelse{\boolean{articletitles}}{{\it {The \lhcb
  simulation application, \gauss: design, evolution and experience}},
  }{}\href{http://dx.doi.org/10.1088/1742-6596/331/3/032023}{{J.\ Phys.\ Conf.\
  Ser.\ } {\bf 331} (2011) 032023}\relax
\mciteBstWouldAddEndPuncttrue
\mciteSetBstMidEndSepPunct{\mcitedefaultmidpunct}
{\mcitedefaultendpunct}{\mcitedefaultseppunct}\relax
\EndOfBibitem
\bibitem{LHCb-PAPER-2012-035}
LHCb collaboration, R.~Aaij {\em et~al.},
  \ifthenelse{\boolean{articletitles}}{{\it {Measurement of the time-dependent
  $CP$ asymmetry in $B^0 \to \jpsi K^0_S$ decays}},
  }{}\href{http://dx.doi.org/10.1016/j.physletb.2013.02.054}{Phys.\ Lett.\
  {\bf B721} (2013) 24}, \href{http://arxiv.org/abs/1211.6093}{{\tt
  arXiv:1211.6093}}\relax
\mciteBstWouldAddEndPuncttrue
\mciteSetBstMidEndSepPunct{\mcitedefaultmidpunct}
{\mcitedefaultendpunct}{\mcitedefaultseppunct}\relax
\EndOfBibitem
\bibitem{LHCb-PAPER-2013-015}
LHCb collaboration, R.~Aaij {\em et~al.},
  \ifthenelse{\boolean{articletitles}}{{\it {Measurement of the effective
  $B^0_s \to J/\psi K^0_S$ lifetime}},
  }{}\href{http://dx.doi.org/10.1016/j.nuclphysb.2013.04.021}{Nucl.\ Phys.\
  {\bf B873} (2013) 275}, \href{http://arxiv.org/abs/1304.4500}{{\tt
  arXiv:1304.4500}}\relax
\mciteBstWouldAddEndPuncttrue
\mciteSetBstMidEndSepPunct{\mcitedefaultmidpunct}
{\mcitedefaultendpunct}{\mcitedefaultseppunct}\relax
\EndOfBibitem
\bibitem{LHCb-PAPER-2013-061}
LHCb collaboration, R.~Aaij {\em et~al.},
  \ifthenelse{\boolean{articletitles}}{{\it {Searches for $\Lambda^0_{b}$ and
  $\Xi^0_b$ decays to $K^0_S p \pi^-$ and $K^0_S p K^-$ final states with first
  observation of the $\Lambda^0_b \to K^0_S p \pi^-$ decay}},
  }{}\href{http://dx.doi.org/10.1007/JHEP04(2014)087}{JHEP {\bf 04} (2014)
  087}, \href{http://arxiv.org/abs/1402.0770}{{\tt arXiv:1402.0770}}\relax
\mciteBstWouldAddEndPuncttrue
\mciteSetBstMidEndSepPunct{\mcitedefaultmidpunct}
{\mcitedefaultendpunct}{\mcitedefaultseppunct}\relax
\EndOfBibitem
\bibitem{LHCb-PAPER-2014-006}
LHCb collaboration, R.~Aaij {\em et~al.},
  \ifthenelse{\boolean{articletitles}}{{\it {Differential branching fractions
  and isospin asymmetries of $B \to K^{(*)}\mu^+\mu^-$ decays}},
  }{}\href{http://dx.doi.org/10.1007/JHEP06(2014)133}{JHEP {\bf 06} (2014)
  133}, \href{http://arxiv.org/abs/1403.8044}{{\tt arXiv:1403.8044}}\relax
\mciteBstWouldAddEndPuncttrue
\mciteSetBstMidEndSepPunct{\mcitedefaultmidpunct}
{\mcitedefaultendpunct}{\mcitedefaultseppunct}\relax
\EndOfBibitem
\bibitem{LHCb-PAPER-2014-016}
LHCb collaboration, R.~Aaij {\em et~al.},
  \ifthenelse{\boolean{articletitles}}{{\it {Observation of the $B^0_s\to
  J/\psi K_{{\rm S}}^0 K^\pm \pi^\mp$ decay}},
  }{}\href{http://dx.doi.org/10.1007/JHEP07(2014)140}{JHEP {\bf 07} (2014)
  140}, \href{http://arxiv.org/abs/1405.3219}{{\tt arXiv:1405.3219}}\relax
\mciteBstWouldAddEndPuncttrue
\mciteSetBstMidEndSepPunct{\mcitedefaultmidpunct}
{\mcitedefaultendpunct}{\mcitedefaultseppunct}\relax
\EndOfBibitem
\bibitem{PDG2012}
Particle Data Group, J.~Beringer {\em et~al.},
  \ifthenelse{\boolean{articletitles}}{{\it {\href{http://pdg.lbl.gov/}{Review
  of particle physics}}},
  }{}\href{http://dx.doi.org/10.1103/PhysRevD.86.010001}{Phys.\ Rev.\  {\bf
  D86} (2012) 010001}, {and 2013 partial update for the 2014 edition}\relax
\mciteBstWouldAddEndPuncttrue
\mciteSetBstMidEndSepPunct{\mcitedefaultmidpunct}
{\mcitedefaultendpunct}{\mcitedefaultseppunct}\relax
\EndOfBibitem
\bibitem{Feindt:2006pm}
M.~Feindt and U.~Kerzel, \ifthenelse{\boolean{articletitles}}{{\it {The
  NeuroBayes neural network package}},
  }{}\href{http://dx.doi.org/10.1016/j.nima.2005.11.166}{Nucl.\ Instrum.\
  Meth.\  {\bf A559} (2006) 190}\relax
\mciteBstWouldAddEndPuncttrue
\mciteSetBstMidEndSepPunct{\mcitedefaultmidpunct}
{\mcitedefaultendpunct}{\mcitedefaultseppunct}\relax
\EndOfBibitem
\bibitem{Pivk:2004ty}
M.~Pivk and F.~R. Le~Diberder, \ifthenelse{\boolean{articletitles}}{{\it
  {sPlot: a statistical tool to unfold data distributions}},
  }{}\href{http://dx.doi.org/10.1016/j.nima.2005.08.106}{Nucl.\ Instrum.\
  Meth.\  {\bf A555} (2005) 356},
  \href{http://arxiv.org/abs/physics/0402083}{{\tt
  arXiv:physics/0402083}}\relax
\mciteBstWouldAddEndPuncttrue
\mciteSetBstMidEndSepPunct{\mcitedefaultmidpunct}
{\mcitedefaultendpunct}{\mcitedefaultseppunct}\relax
\EndOfBibitem
\bibitem{Punzi:2003bu}
G.~Punzi, \ifthenelse{\boolean{articletitles}}{{\it {Sensitivity of searches
  for new signals and its optimization}}, }{} in {\em Statistical Problems in
  Particle Physics, Astrophysics, and Cosmology} (L.~{Lyons}, R.~{Mount}, and
  R.~{Reitmeyer}, eds.), p.~79, 2003.
\newblock \href{http://arxiv.org/abs/physics/0308063}{{\tt
  arXiv:physics/0308063}}\relax
\mciteBstWouldAddEndPuncttrue
\mciteSetBstMidEndSepPunct{\mcitedefaultmidpunct}
{\mcitedefaultendpunct}{\mcitedefaultseppunct}\relax
\EndOfBibitem
\bibitem{Skwarnicki:1986xj}
T.~Skwarnicki, {\em {A study of the radiative cascade transitions between the
  Upsilon-prime and Upsilon resonances}}, PhD thesis, Institute of Nuclear
  Physics, Krakow, 1986,
  {\href{http://inspirehep.net/record/230779/files/230779.pdf}{DESY-F31-86-02}}\relax
\mciteBstWouldAddEndPuncttrue
\mciteSetBstMidEndSepPunct{\mcitedefaultmidpunct}
{\mcitedefaultendpunct}{\mcitedefaultseppunct}\relax
\EndOfBibitem
\bibitem{Cranmer:2000du}
K.~S. Cranmer, \ifthenelse{\boolean{articletitles}}{{\it {Kernel estimation in
  high-energy physics}},
  }{}\href{http://dx.doi.org/10.1016/S0010-4655(00)00243-5}{Comput.\ Phys.\
  Commun.\  {\bf 136} (2001) 198},
  \href{http://arxiv.org/abs/hep-ex/0011057}{{\tt arXiv:hep-ex/0011057}}\relax
\mciteBstWouldAddEndPuncttrue
\mciteSetBstMidEndSepPunct{\mcitedefaultmidpunct}
{\mcitedefaultendpunct}{\mcitedefaultseppunct}\relax
\EndOfBibitem
\bibitem{LHCb-PAPER-2013-048}
LHCb collaboration, R.~Aaij {\em et~al.},
  \ifthenelse{\boolean{articletitles}}{{\it {Measurement of the charge
  asymmetry in $B^\pm \to \phi K^\pm$ and search for $B^\pm \to \phi \pi^\pm$
  decays}}, }{}\href{http://dx.doi.org/10.1016/j.physletb.2013.11.036}{Phys.\
  Lett.\  {\bf B728} (2014) 85}, \href{http://arxiv.org/abs/1309.3742}{{\tt
  arXiv:1309.3742}}\relax
\mciteBstWouldAddEndPuncttrue
\mciteSetBstMidEndSepPunct{\mcitedefaultmidpunct}
{\mcitedefaultendpunct}{\mcitedefaultseppunct}\relax
\EndOfBibitem
\bibitem{Cheng:2009mu}
H.-Y. Cheng and C.-K. Chua, \ifthenelse{\boolean{articletitles}}{{\it {QCD
  factorization for charmless hadronic \Bs decays revisited}},
  }{}\href{http://dx.doi.org/10.1103/PhysRevD.80.114026}{Phys.\ Rev.\  {\bf
  D80} (2009) 114026}, \href{http://arxiv.org/abs/0910.5237}{{\tt
  arXiv:0910.5237}}\relax
\mciteBstWouldAddEndPuncttrue
\mciteSetBstMidEndSepPunct{\mcitedefaultmidpunct}
{\mcitedefaultendpunct}{\mcitedefaultseppunct}\relax
\EndOfBibitem
\bibitem{Ali:2007ff}
A.~Ali {\em et~al.}, \ifthenelse{\boolean{articletitles}}{{\it {Charmless
  non-leptonic $\Bs$ decays to $PP$, $PV$ and $VV$ final states in the pQCD
  approach}}, }{}\href{http://dx.doi.org/10.1103/PhysRevD.76.074018}{Phys.\
  Rev.\  {\bf D76} (2007) 074018},
  \href{http://arxiv.org/abs/hep-ph/0703162}{{\tt arXiv:hep-ph/0703162}}\relax
\mciteBstWouldAddEndPuncttrue
\mciteSetBstMidEndSepPunct{\mcitedefaultmidpunct}
{\mcitedefaultendpunct}{\mcitedefaultseppunct}\relax
\EndOfBibitem
\bibitem{Su:2011eq}
F.~Su, Y.-L. Wu, C.~Zhuang, and Y.-B. Yang,
  \ifthenelse{\boolean{articletitles}}{{\it {Charmless $\Bs\to PP, PV, VV$
  decays based on the six-quark effective Hamiltonian with strong phase effects
  II}}, }{}\href{http://dx.doi.org/10.1140/epjc/s10052-012-1914-4}{Eur.\ Phys.\
  J.\  {\bf C72} (2012) 1914}, \href{http://arxiv.org/abs/1107.0136}{{\tt
  arXiv:1107.0136}}\relax
\mciteBstWouldAddEndPuncttrue
\mciteSetBstMidEndSepPunct{\mcitedefaultmidpunct}
{\mcitedefaultendpunct}{\mcitedefaultseppunct}\relax
\EndOfBibitem
\end{mcitethebibliography}

\newpage
\centerline{\large\bf LHCb collaboration}
\begin{flushleft}
\small
R.~Aaij$^{41}$, 
B.~Adeva$^{37}$, 
M.~Adinolfi$^{46}$, 
A.~Affolder$^{52}$, 
Z.~Ajaltouni$^{5}$, 
S.~Akar$^{6}$, 
J.~Albrecht$^{9}$, 
F.~Alessio$^{38}$, 
M.~Alexander$^{51}$, 
S.~Ali$^{41}$
% ...

{\footnotesize \it
% ...
$ ^{5}$Clermont Universit\'{e}, Universit\'{e} Blaise Pascal, CNRS/IN2P3, LPC, Clermont-Ferrand, France\\
$ ^{6}$CPPM, Aix-Marseille Universit\'{e}, CNRS/IN2P3, Marseille, France\\
% ...
$ ^{9}$Fakult\"{a}t Physik, Technische Universit\"{a}t Dortmund, Dortmund, Germany\\
% ...
$ ^{37}$Universidad de Santiago de Compostela, Santiago de Compostela, Spain\\
$ ^{38}$European Organization for Nuclear Research (CERN), Geneva, Switzerland\\
% ...
$ ^{41}$Nikhef National Institute for Subatomic Physics, Amsterdam, The Netherlands\\
% ...
$ ^{46}$H.H. Wills Physics Laboratory, University of Bristol, Bristol, United Kingdom\\
% ...
$ ^{51}$School of Physics and Astronomy, University of Glasgow, Glasgow, United Kingdom\\
$ ^{52}$Oliver Lodge Laboratory, University of Liverpool, Liverpool, United Kingdom\\
% ...
}
\end{flushleft}

\end{document}